\documentclass[apjl]{emulateapj}
\usepackage{graphicx}
\usepackage{natbib}
\usepackage{amsmath}
\usepackage{multirow}
\usepackage{array}
\usepackage{subfigure}
\usepackage{float}

\begin{document}

\title{Evolutionary Models of Super-Earths and Mini-Neptunes Incorporating Cooling and Mass Loss}
\author{Alex R. Howe}
\affil{Department of Astrophysical Sciences, Princeton University}
\affil{Peyton Hall, Princeton, NJ 08544, USA}
\affil{arhowe@astro.princeton.edu}
\author{Adam Burrows}
\affil{Department of Astrophysical Sciences, Princeton University}
\affil{Peyton Hall, Princeton, NJ 08544, USA}
\affil{burrows@astro.princeton.edu}


\begin{abstract}

We construct models of the structural evolution of super-Earth- and mini-Neptune-type exoplanets with H$_2$-He envelopes, incorporating radiative cooling and XUV-driven mass loss. We conduct a parameter study of these models, focusing on initial mass, radius, and envelope mass fractions, as well as orbital distance, metallicity, and the specific prescription for mass loss. From these calculations, we investigate how the observed masses and radii of exoplanets today relate to the distribution of their initial conditions. Orbital distance and initial envelope mass fraction are the most important factors determining planetary evolution, particular radius evolution. Initial mass also becomes important below a ``turnoff mass,'' which varies with orbital distance, with mass-radius curves being approximately flat for higher masses. Initial radius is the least important parameter we study, with very little difference between the hot start and cold start limits after an age of 100 Myr. Model sets with no mass loss fail to produce results consistent with observations, but a plausible range of mass loss scenarios is allowed. In addition, we present scenarios for the formation of the Kepler-11 planets. Our best fit to observations Kepler-11b and Kepler-11c involves formation beyond the snow line, after which they moved inward, circularized, and underwent a reduced degree mass loss.

\end{abstract}

\maketitle


\section{Introduction}
\label{intro}

Recent exoplanet surveys, particularly the {\it Kepler} Mission, have discovered hundreds of confirmed exoplanets and thousands of candidates, prompting studies of the general structures and evolution of planetary bodies. Notably, the most common class of planets discovered by these surveys are ``super-Earths'' or ``mini-Neptunes'' with radii of $\sim$1-4 R$_\Earth$, with an apparent modal value of $\sim$2-3 R$_\Earth$, for which there is no analog in our Solar System \citep{2014ApJS..210...19B}. This appears to be true even for a sample set of planets corrected for completeness \citep{2014arXiv1406.6048S}. For objects for which masses have been measured, the inferred densities of planets with radii of 2-4 R$_\Earth$ are too low to be accounted for without the presence of a voluminous hydrogen-helium envelope. However, their densities fall in a wide range so that planets of the same mass may have very different radii; they may be terrestrial, or they may have envelopes comprising up to tens of percent of their mass \citep{2014ApJS..210...20M,2014ApJ...787..173H}.

A small number of very-low-density planets have been observed with radii $>$5 R$_\Earth$, but masses less than that of Neptune, such as Kepler-18c and -18d \citep{2011ApJS..197....7C}, Kepler-51b \citep{2014ApJ...783...53M}, Kepler-79d \citep{2014ApJ...785...15J}, Kepler-87b \citep{2014A&A...561A.103O}, and Kepler-89e \citep{2013ApJ...778..185M}, as demonstrated variously by radial velocity (RV) and transit timing variation (TTV) measurements. These objects have so far been found only at larger orbital distances of $~$0.1-0.2 AU, which, along with theoretical considerations, suggests that they typically do not retain a H$_2$-He atmosphere at $<$0.1 AU due to rapid mass loss or other factors.

Many of these planets lie within 0.2 AU of their host stars, suggesting at the very least that they may have been significantly sculpted by stellar flux-driven mass loss. This is true even at larger masses up to the ``hot Jupiters''; for example, HD 209458b appears to be experiencing measurable mass loss today on the order of $2\times 10^{10}$ g s$^{-1}$ ($10^{-10}$ M$_\Earth$ yr$^{-1}$) \citep{2008A&A...483..933E}, and some apparently-rocky planets may, in fact, have formed with significant envelopes that have since been entirely lost. Therefore, a self-consistent model of extrasolar planets with hydrogen-helium atmospheres should incorporate the effects of Kelvin-Helmholtz cooling and mass loss to determine whether a particular planet has likely been subjected to significant alteration by mass loss.

Structural models of planets indicate that a planet's radius may serve as a reasonably good proxy for estimating its composition$-$either the envelope mass fraction \citep{2014ApJ...792....1L} or the mass of the envelope itself \citep{2014ApJ...787..173H}. However, there are many degeneracies involved in these estimates, and uncertainties in the data cause further difficulties in interpretation. However, combining good mass, radius, and age measurements with evolutionary models may allow less ambiguous interpretations to be made.

An evolutionary history of a planet incorporating the effects of radiative cooling and mass loss (particularly mass loss due to XUV photoevaporation) could also give clues to planet formation. \citet{2005ApJ...622..680R} found that solar-type stars produce the greatest amount of ultraviolet radiation when they are young, declining rapidly after 100 Myr. Therefore, if such an evolutionary model can determine whether a given planet could have survived in its present location during the high-mass-loss early phases of its existence, it would provide further information about where and when planets form in their proto-planetary disks.

There is some uncertainty in the effect of mass loss. \citet{2012ApJ...761...59L} constructed evolutionary models incorporating cooling and mass loss for the specific cases of the planets of the Kepler-11 system. They specifically adopted an energy-limited XUV-driven mass loss prescription, finding that the five nearest planets to the star (within 0.25 AU) have undergone significant mass loss, but that only the innermost has potentially lost its entire envelope. However, \citet{2013ApJ...776....2L}, in a similar study of the Kepler-36 system, found that the transition in orbital distance from retaining nearly all of the envelope to losing all or nearly all of it is relatively sharp for a given core mass.

Previously, in \citet{2014ApJ...787..173H}, we studied the structures of exoplanets with gaseous envelopes, finding that observed extrasolar planets may have a wide range of core masses and envelope masses with little correlation between the two. Therefore, a wide range of models is needed to fully characterize the exoplanet population. For evolutionary models, this problem is compounded because the planets may have a wide range of orbital distances and orbit stars with a range of spectral types, resulting in a wide range of temperatures of their atmospheres. Also, the initial radii of the planets are not certain, although some work has been done to model them, e.g. \citet{2012A&A...547A.112M,2014ApJ...795...65J}. Finally, the detailed properties of the atmospheres, such as cloud cover, redistribution of heat, and the processes that contribute to mass loss, are poorly understood.

While there is a rich history of exoplanet structure modeling and a growing body of work on planetary evolution, much remains to be done. Uncertainties must be addressed and plausible parameters set to produce useful models. Ideally, a range of models would be produced with different parameters, but in the case of atmospheric properties, which are not fully understood, a simple parameterization is perforce necessary. Also, these techniques to study evolution have only recently been applied to super-Earths and mini-Neptunes \citep{2012ApJ...761...59L}.

Our paper investigates the evolution of theoretical super-Earth and mini-Neptune models in different planetary systems over billions of years. We focus particularly on finding the boundaries of the parameter space in which a planet's envelope is lost completely and on estimating the initial conditions of observed planets.

Section \ref{previous} summarizes the previous work on this topic. Section \ref{model} describes our planetary model. We demonstrate our code in Section \ref{structures} by providing internal density and pressure profiles over time for example models. Then, in Section \ref{mass_loss}, we show the resultant mass loss rates and investigate the effects of varying the prescription for mass loss. In Section  \ref{parameter_study}, we conduct a parameter study of the most important physical parameters that go into the calculations, constructing models with a range of initial masses, radii, envelope mass fractions, orbital distances, metallicities, and mass loss prescriptions. In Section \ref{kepler-11}, we attempt to fit the evolutionary history of the planets in the Kepler-11 system. Section \ref{transit} provides example analyses of the evolution of transit spectra of our models, and we present general conclusions in Section \ref{conclusions}.


\section{Previous Work on Evolutionary Models and Atmospheric Escape}
\label{previous}

One important early study of atmospheric escape was undertaken by \citet{2007Icar..187..358H}. They estimated mass loss by scaling the XUV flux model of \citet{1981Icar...48..150W} to apply to highly-irradiated giant planets. They applied the tidal effect to both the Watson models and the model of \citet{2004A&A...419L..13B}. They also added an efficiency parameter$-$the fraction of the total stellar flux that contributes to energy-limited escape, which is less than the total XUV flux due to losses through dissociation and ionization, in contrast with the Baraffe model, which used the full XUV flux. Further, they applied the XUV flux model of \citet{2005ApJ...622..680R}. The overall efficiency parameter (the fraction of total stellar flux that goes into the evaporation of the atmosphere) at 4.5 Gyr is $\sim 10^{-4}$ for the Baraffe model, compared with $\sim 10^{-6}$ for the Watson model. The efficiency parameter is proportional to the XUV flux, which is considered to decrease at a rate proportional to $t^{-1.23}$ after 100 Myr. Therefore, \citet{2007Icar..187..358H} found a mass loss rate a factor of 100 smaller than \citet{2004A&A...419L..13B}, which eliminates the problem of improbable model parameters for HD 209458b found by that study. They found that mass loss becomes significant for highly-irradiated planets in general only for planets of about the mass of Saturn and smaller.

\citet{2012A&A...547A.112M} and \citet{2012A&A...547A.111M} modeled a wide range of planets with radioactive heating and Kelvin-Helmholtz cooling, but not mass loss. Notably, they also modeled planet formation with a core-accretion model incorporating migration, providing estimates for initial radii at the end of the accretion phase, and modeled radiative cooling with gray atmospheric boundary conditions in the Eddington approximation. However, they set a minimum orbital distance of 0.1 AU (as opposed to an observed limit of about 0.015 AU for 55 Cancri e) so that the strongest irradiation effects are avoided. Using this framework, they model the formation and evolution of planets with masses from 1 to 10,000 M$_\Earth$ (0.003 to 30 $M_J$) and envelope mass fractions ($f_{env}$) from $<1\%$ to $>99\%$. These objects have radii of $\lesssim 20 R_\Earth$ ($\lesssim 2 R_J$) once the accretion phase ends and follow a relatively narrow mass-radius relation after 100 Myr for which radius is a good first-order proxy for the fraction of mass in the envelope, $f_{env}$ and mass and radius together are sufficient to compute $f_{env}$ with relatively high precision. However, in all cases, they assumed a low-entropy ``cold start'' scenario in which the energy of accretion is thermally radiated on a timescale that is short compared with the accretion timescale, so that the effective temperature is always relatively low. This is in contrast to a ``hot start,'' in which the energy of accretion is not radiated efficiently, resulting in a much greater envelope entropy and effective temperature when accretion ends.

Meanwhile, \citet{2009ApJ...693...23M} modeled mass loss rates for hot Jupiters as a function of XUV flux, taking into account additional processes such as ionization, radiative losses, plasma recombination, tidal gravity, pressure confinement by the stellar wind, and magnetic pressure. They found that mass loss rates for hot Jupiters are approximately energy-limited for lower fluxes ($<10^4$ erg cm$^{-2}$ s$^{-1}$ or about the flux received by a planet 0.2 AU away from a young G dwarf at $<$100 Myr in the Watson model). The nearly energy-limited mass loss follows a relation of $\dot{m} \propto F_{XUV}^{0.9}$. However, at higher fluxes, this relation becomes flatter, roughly $\dot{m} \propto F_{XUV}^{0.6}$ due to radiative and recombination losses.

\citet{2013ApJ...775..105O} examined a detailed model of mass loss going beyond the simple, energy-limited prescription. They incorporated X-ray emissions in addition to XUV, and they accounted for the recombination and line cooling processes that can reduce the mass loss efficiency. They modeled both Jupiter-mass and Neptune-mass planets based on a structure model using the {\sc mesa} code with a range of initial radii from about 10 to more than 20 R$_\Earth$$-$notably including Neptune-mass models with large cores of 12-15 M$_\Earth$. With this mass loss prescription, they found significant mass loss approaching most of the hydrogen fraction (and a consummate decrease in radius) for highly irradiated planets closer than $\sim$0.1 AU.

\citet{2014ApJ...795...65J} added atmospheric escape to the population synthesis model of \citet{2012A&A...547A.112M} and \citet{2012A&A...547A.111M}, using multiple mass loss prescriptions, including XUV-driven mass loss only, EUV plus X-ray-driven mass loss, purely energy-limited mass loss, and mass loss that is moderated by cooling radiation in the high-EUV regime, in which case the mass outflow is at the sound speed. In order to better study evaporation, they shifted the entire planet population inward by 0.04 AU, placing the inner edge at 0.06 AU.

\citet{2012ApJ...761...59L} modeled thermal evolution and XUV-driven mass loss for low-mass, super-Earth/mini-Neptune planets, rather than giant planets. In these models, a planetary structure is modeled with a fully convective envelope, with a structure determined by the specific entropy, which cools radiatively over time. They incorporated radioactive heating and the thermal inertia of a rocky core. The radiative flux is determined from a grid of model atmospheres based on the models of \citet{2011ApJ...733....2N}. They used a ``hot start'' with a large initial entropy. The XUV-driven mass loss in this model is computed by an energy-limited formula with an efficiency of $10^{+10}_{-5}\%$ applied to the XUV flux specifically and an XUV-photosphere at a pressure level from 0.1 nbar to 10 nbar. A more precise calculation from the \citet{2005ApJ...622..680R} results in an efficiency parameter under the \citet{2007Icar..187..358H} formulation of $3.3\times 10^{-6}$ at 4.5 Gyr.

\citet{2012ApJ...761...59L} apply their models to the super-Earths and mini-Neptunes, Kepler-11b through f, in an attempt to determine their initial compositions. In their best fits, all five planets formed with hydrogen-helium envelopes comprising at least a few percent of their masses, and they all subsequently lost at least 20\% of the envelope masses after 100 Myr, with Kepler-11b being almost entirely stripped.

In \citet{2013ApJ...776....2L}, these same models were applied to Kepler-36b and -36c. Kepler-36b (mass 4.5 M$_\Earth$) appears to be entirely rocky, while Kepler-36c (mass 8.1 M$_\Earth$) has a density less than that of water, suggesting a voluminous hydrogen-helium atmosphere, despite being in orbits at similar distances (an apparent 6:7 mean motion resonance). They model mass loss over 5 Gyr and find that the density dichotomy can be explained by a 65\% more massive core for Kepler-36c, which results in a mass loss timescale $\sim$3 times as long as for Kepler-36b, allowing it to retain an envelope over its lifetime.

\citet{2014ApJ...792....1L} apply these models to the general case, producing models with a wide range of masses (1-20 M$_\Earth$), envelope mass fractions, $f_{env}$ (0.01\%-20\%), and levels of stellar irradiation (0.1-1000 times Earth's irradiation, F$_\Earth$). They find that, to first order, planetary radius is dependent on only $f_{env}$ for larger envelopes of $f_{env}>1\%$. They also propose a dividing line of 1.75 R$_\Earth$ between planets that are entirely rocky and those with gaseous envelopes.

\citet{2014arXiv1409.2982W} apply the models of \citet{2012ApJ...761...59L} and subsequent papers to the observed distribution of planetary radii from the {\it Kepler} mission (selected for completion) using a hierarchical Bayesian model. They find that the observed population has a distribution of $f_{env}$ with a mean of 0.7\% and a standard deviation of 0.6 dex, corresponding to a modal radius of $\sim$2.0-2.5 R$_\Earth$.

\citet{2014ApJ...783L...6W} fit a mass-radius function to 65 exoplanets with radii less than 4 R$_\Earth$. In doing so, they set a boundary of 1.5 R$_\Earth$ between rocky planets and those with gaseous envelopes. For rocky planets their best fit is a density of $\rho = 2.43 + 3.39(R/R_\Earth)$ g cm$^{-3}$, while for planets with gaseous envelopes, their fit for the average radius is a mass of $(M/M_\Earth) = 2.69(R/R_\Earth)^{0.93}$. Their mass, radius, and density measurements are taken from \citet{2014ApJS..210...20M}.

\begin{figure}[htp]
\includegraphics[width=\columnwidth]{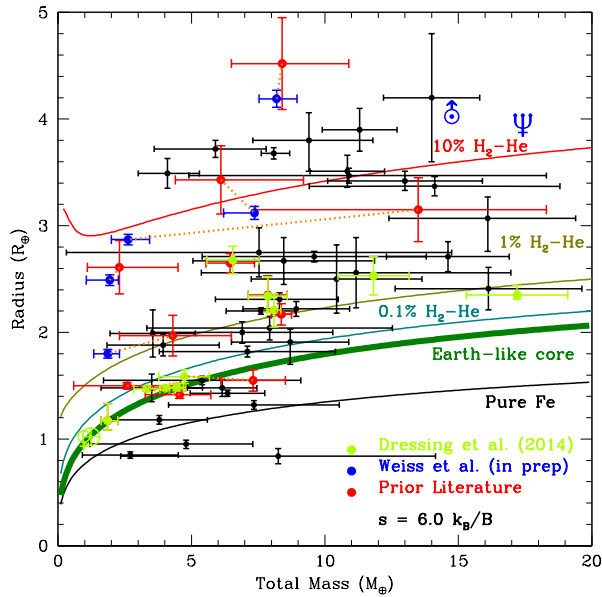}
\caption{Mass-radius diagram for low-mass exoplanets with observed masses and radii, compared with mass-radius curves for selected terrestrial and gaseous compositions with an envelope entropy of 6.0 $k_B/B$. Based on data from \citet{2014ApJ...787..173H}.}
\label{mass-rad}
\end{figure}

In our previous work, \citet{2014ApJ...787..173H}, we presented structural models of super-Earths and mini-Neptunes without thermal evolution, but at a range of envelope entropies, potentially representing different ages or levels of irradiation. In Figure \ref{mass-rad}, we plot mass-radius curves for a few selected compositions with observed mass and radius measurements of planets. This figure includes models with pure iron and terrestrial compositions plus terrestrial\footnote{A core-mantle structure of 32.5\% Fe and 67.5\% MgSiO$_3$.} compositions with H$_2$-He envelopes comprising 0.1\%, 1\%, and 10\% of their masses with entropies of 6.0 $k_B$ per baryon. We also found that radius is a proxy for composition, but proposed that it is more dependent on the envelope mass, $m_{env}$. On the other hand, planets with similar masses may vary greatly in radius. Assuming an Earth-like core, the range of observed radii may indicate envelope masses from 0.01 $M_\Earth$ to several $M_\Earth$ for a given total mass between 2 and 10 $M_\Earth$ ($f_{env}$ on the order of 0.1\% to 50\%).

\begin{figure}[htp]
\includegraphics[width=\columnwidth]{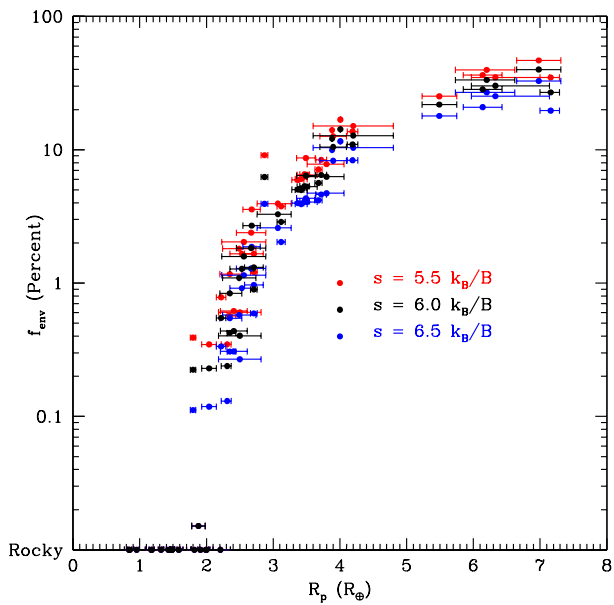}
\caption{Our calculated envelope fractions (based on numerical data from \citealp{2014ApJ...787..173H}) for planets with observed masses and radii versus radius. We compute models with envelope entropies of 5.5, 6.0, and 6.5 $k_b$ per baryon.}
\label{planets}
\end{figure}

These results are updated and summarized in Figure \ref{planets} for the case of an Earth-like composition core and an envelope including a radiative layer. Figure \ref{planets} shows the relationship between envelope fraction and planetary radius for planets with observed masses and radii for envelope entropies of 5.5, 6.0, and 6.5 $k_b$ per baryon. These entropies are representative of those found by \citet{2014ApJ...792....1L}. We consider ``rocky'' or terrestrial planets to be those with a convective envelope with a mass fraction $f_{env}<0.0001$, excluding the (much lower mass) radiative atmosphere. Often, a bare core will fit the observed radius. We find that the values are closely correlated, with radius increasing rapidly for high envelope fractions. This result is qualitatively similar to that of \citet{2014arXiv1409.2982W}, but with greater scatter and a higher apparent median value of $f_{env}$ of several percent as opposed to 0.7\%. However, this observational sample is not necessarily complete. We also find that the lowest-density planets with radii larger than Neptune have $f_{env}>10\%$. (Compare Uranus and Neptune, which have $f_{env}$ = 10\%-20\%, \citealp{2011ApJ...726...15H})

\citet{2014ApJ...787..173H} also presented fits with models with pure water ice cores, which produced similar results, except that the envelope masses were smaller by tens of percent; for small planets ($R\lesssim 2\,R_\Earth$), no envelope was needed to fit the data. Some additional uncertainty remained because a range of envelope entropies is possible. We found that a range of 5.5-6.5 $k_B$ per baryon in entropy was associated with an uncertainty of tens of percent in envelope masses, roughly on the same order as the observational uncertainties.


\section{Evaporation and Evolution Model}
\label{model}

We compute planetary structure profiles using the code we describe in \citet{2014ApJ...787..173H}. In order to model thermal evolution, we add to the equations of hydrostatic equilibrium,
\begin{eqnarray}
\frac{dP}{dr}=-\frac{Gm\rho}{r^2} \\
\frac{dm}{dr}=4\pi r^2\rho,
\end{eqnarray}
equations of thermal transport and energy production:
\begin{eqnarray}
\frac{dT}{dr}=\left(1-\frac{1}{\gamma}\right)\frac{T}{P}\frac{dP}{dr} \\
\frac{dL}{dr}=4\pi r^2\rho\varepsilon,
\end{eqnarray}
where $T$ is the temperature, $L$ is the luminosity, $\gamma$ is the adiabatic index, $\varepsilon$ is the energy production per unit mass, e.g. from radioactive decay. A convective thermal gradient in the atmosphere is assumed.

We model radiative cooling by relating the energy budget to the change in envelope entropy:
\begin{equation}
\int_{M_{core}}^{M_p}T\frac{dS}{dt}dm = -L_{int} + L_{radio} - c_vM_{core}\frac{dT_{core}}{dt},
\end{equation}
where the integral is over the mass of the H$_2$-He envelope.

$L_{int}$ is termed the ``intrinsic'' luminosity of the planet, that is, the net luminosity of cooling, equal to the total luminosity minus the ``external'' luminosity$-$the intercepted stellar flux$-$so that $L_{int} = L_{tot}-L_{ext}$. Each of these components may be associated with a temperature, most importantly, the {\it effective temperature}, defined by $L_{int} = 4\pi R^2\sigma T_{eff}^4$.

Energy is added to the envelope by radioactive decay, $L_{radio}$, and the cooling of the core, $c_vM_{core}\frac{dT_{core}}{dt}$, which we model as an isothermal ``hot rock,'' where $c_v$ is the specific heat capacity. In this paper, unless otherwise specified, we define the core (or the composition of a purely ``terrestrial'' planet) as having an Earth-like core-mantle structure of 32.5\% Fe and 67.5\% MgSiO$_3$.

The convective portion of the envelope is modeled with an equation of state for H$_2$-He, as described in \citet{2014ApJ...787..173H}, with metals included as an ideal gas with the volume addition law. We include upper and lower boundary conditions to determine the structure of the envelope. The lower boundary condition is simply that the base of the envelope must be continuous in temperature and pressure with an isothermal core. For the upper boundary condition, we compute a grid of atmosphere models using the CoolTLUSTY code \citep{2003ApJ...594.1011H,2006ApJ...650.1140B}, incorporating stellar irradiation, assuming complete redistribution and reradiation of the stellar flux over the surface. Interpolating in this grid provides a relation between effective temperature, entropy, and gravity.

We decree that the base of the radiative atmosphere must be continuous in pressure and entropy with the isentropic envelope and use the entropy and gravity to determine the effective temperature at the radiative-convective boundary that we use to compute the cooling rate of the model. The radiative atmosphere also adds an additional amount, $\Delta R$, to the radius of the planet computed by the structural code. This $\Delta R$ may be a significant fraction of the total radius of the planet throughout its lifetime, but especially at early times when the atmosphere is ``puffier''. In all cases, we report the total radius, $R = R_{conv} + \Delta R$. There is a discrepancy between the total radius, which extends up to a pressure level of microbars and the observed transit radius, which is equivalent to cutting the atmosphere off at a lower altitude corresponding to a pressure of tens of millibars (more specifically a vertical Rosseland optical depth of $\tau\sim 0.56$; \citealp{2008A&A...481L..83L}). However, this is a secondary effect on the order of $\sim$0.2 R$_\Earth$ for a representative model in our set, and we find that it has a relatively small effect ($<$10\%) on our fits of initial envelope mass fractions to observed mass and radius data.

For mass loss, we use an energy-limited scheme based on \citet{2011ApJ...733....2N}, which is appropriate for relatively lower irradition levels \citep{2014ApJ...795...65J} and is common in evolutionary modeling:
\begin{eqnarray}
\dot{M}=\frac{\epsilon\pi F_{XUV}R^3_{XUV}}{GM_pK_{tide}} \\
K_{tide}=\left(1-\frac{3}{2\xi}+\frac{1}{2\xi^3}\right) \\
\xi = \frac{R_{Hill}}{R_{XUV}},
\end{eqnarray}
where $F_{XUV}$ is the ionizing flux from the star, and $R_{XUV}$ is the radius at which the atmosphere is optically thick to ionizing radiation, for which we use our total radius, $R = R_{conv} + \Delta R$. $K_{tide}$ is a correction for the tidal distortion by the star of the planet (this is an approximate correction neglecting the dependence of tidal properties on the density profile of the planet), and $\epsilon$ is an efficiency factor that we set by default to $\epsilon=0.1$. Thus, we employ the approximation that a certain fraction of the XUV energy intercepted by the planet goes into stripping the atmosphere. This efficiency may vary under different circumstances and in particular may decrease at high levels of irradiation \citep{2014ApJ...795...65J}, so we compute some models with a lower efficiency and keep this particularly in mind when trying to fit highly irradiated observed objects such as Kepler-11b and 11c (see Section \ref{kepler-11}). We consider the envelope to be lost if $f_{\rm env} < 10^{-4}$.

For the XUV flux history, we use the model of \citet{2005ApJ...622..680R} for Sun-like stars, which gives a flux at 1 AU of
\begin{eqnarray}
F_{XUV} = 504\ {\rm erg\ s^{-1}\ cm^{-2},\ t<100\ Myr} \\
F_{XUV} = 29.7t_9^{-1.23}\ {\rm erg\ s^{-1}\ cm^{-2},\ t>100\ Myr}
\end{eqnarray}
where $t_9$ is the age of the star in Gyr. \citet{2010A&A...511L...8S} show that this XUV flux is approximately the same for stellar types from F7 to M3. However, \citet{2014AJ....148...64S} find a lower initial XUV flux for M dwarfs, so there is still some uncertainty in these models.

We also consider the possibility that planets may form at relatively large distances where mass loss is negligible and migrate in to close orbits where mass loss is high some time after formation, perhaps on the order of 100 Myr. Here, we are proposing migration by secular processes such as planet-planet scattering or the Kozai mechanism that may occur long after formation, followed by tidal circularization. We define a {\it migration time}, $t_M$, before which we set $\epsilon = 0$ (no mass loss) and after which we set $\epsilon > 0$ to simulate this late migration behavior in some models. By default, we set $t_M = 0$, for the case where there is no migration from large distances, or such migration occurs during the planet formation period.

The entropy and composition of the envelope fully determine the structure of the convective region, while for the radiative atmosphere, stellar irradiation must also be taken into account. In practice, to compute an evolutionary track for a planet, we set an initial radius and fit an initial envelope entropy for the appropriate equation of state and atmosphere model.

In principle, the initial radius should be the radius of the planet at the end of the accretion phase of planet formation. However, there are few estimates of ``initial radii'' in the literature. \citet{2014ApJ...797...95L} modeled gas accretion onto solid cores and argued that the outer boundary of the planet during accretion should be the minimum of the Hill radius and the Bondi radius, on the order of $\sim$40 R$_\Earth$ for the planets in question. However, evolutionary models, e.g. \citet{2014ApJ...792....1L}, more commonly set their initial conditions at a ``hot start'' of relatively high entropy or a ``cold start'' of relatively low entropy, resulting in initial radius of $\sim$5-15 R$_\Earth$. While we set the initial radius directly instead of the entropy, we retain this strategy and use an initial radius range of 5-20 R$_\Earth$, in part to ensure speed and accuracy of computation. In Section \ref{init_rad}, we show that for $R_i\gtrsim 10$ R$_\Earth$, the precise initial radius matters very little to the subsequent evolution because the radius contracts very quickly.

For evolution, we consider the envelope to be lost entirely when it reached $f_{env}<0.01\%$ and halt the evolution at that point. We also halt the evolution if the effective temperature of the planet falls below 50 K, which is off our grid of atmosphere models, so a few models terminate before losing their entire envelopes.



\section{Example Internal Structure Profiles}
\label{structures}

\begin{figure}[htp]
\includegraphics[width=\columnwidth]{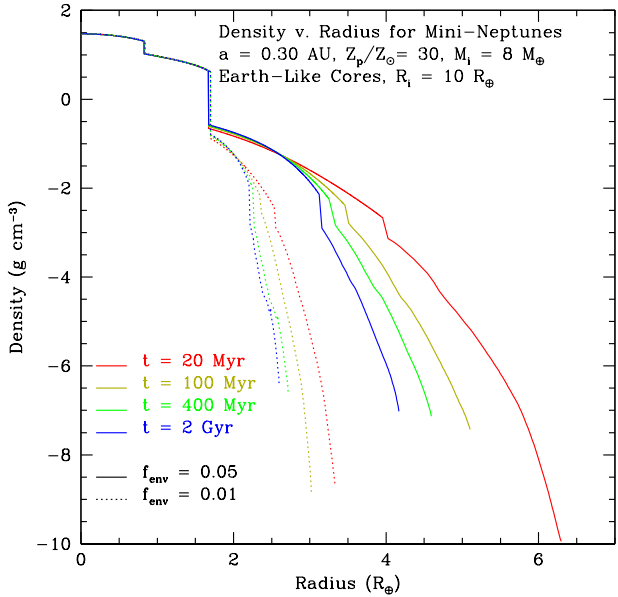}
\caption{Density profiles over time for a planetary model with initial parameters of $M_{tot} = 8$ M$_\Earth$, $f_{env}$ = 0.01 and 0.05, $R_i = 10$ R$_\Earth$, $a = 0.30$ AU, and otherwise using our default parameters.}
\label{density-profile}
\end{figure}

\begin{figure}[htp]
\includegraphics[width=\columnwidth]{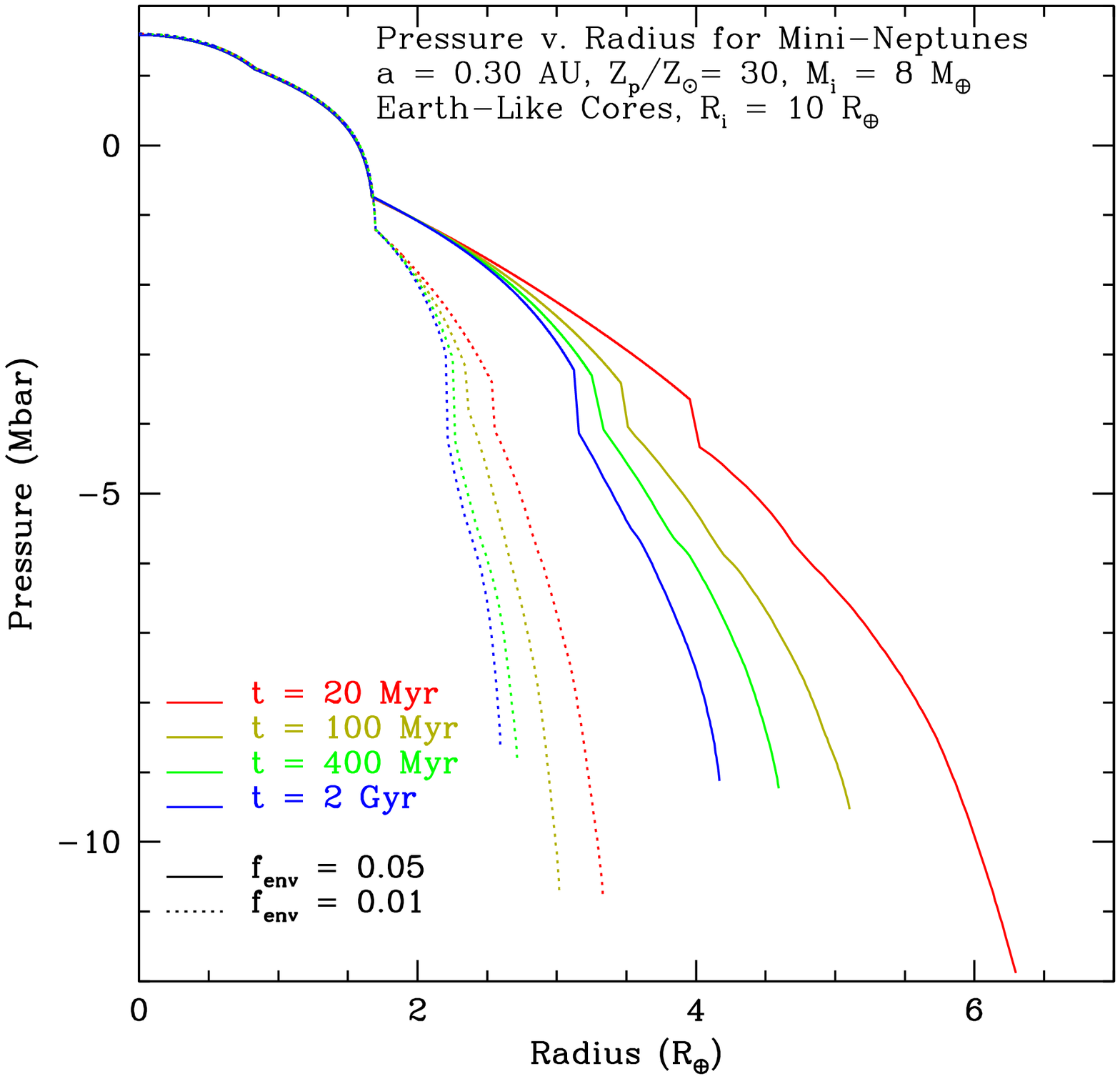}
\caption{Pressure profiles over time for planetary models with initial parameters of $M_{tot} = 8$ M$_\Earth$, $f_{env}$ = 0.01 and 0.05, $R_i = 10$ R$_\Earth$, $a = 0.30$ AU, and otherwise using our default parameters.}
\label{pressure-profile}
\end{figure}

\begin{figure}[htp]
\includegraphics[width=\columnwidth]{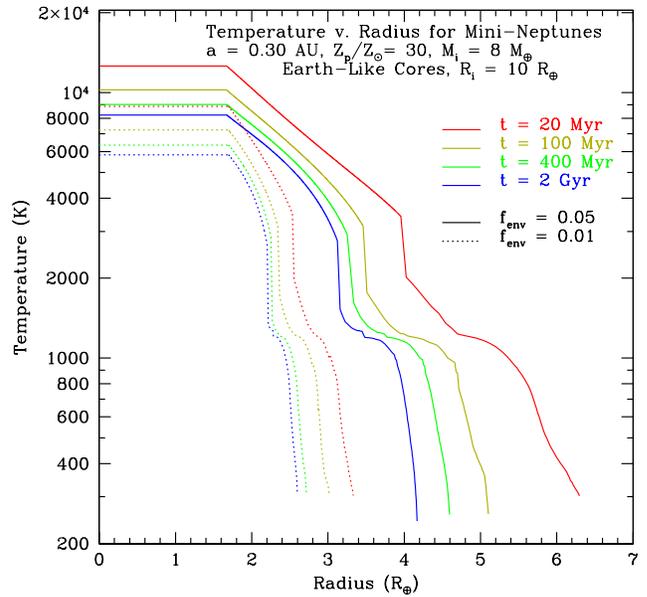}
\caption{Temperature profiles over time for planetary models with initial parameters of $M_{tot} = 8$ M$_\Earth$, $f_{env}$ = 0.01 and 0.05, $R_i = 10$ R$_\Earth$, $a = 0.30$ AU, and otherwise using our default parameters.}
\label{temperature-profile}
\end{figure}

Our code gives us information about the internal structures of our planetary models, such as density, pressure, and temperature profiles of some sample models, and the evolution of those profiles over time. We plot these profiles for two sample models in Figures \ref{density-profile}, \ref{pressure-profile}, and \ref{temperature-profile}, respectively. These models have parameters of initial $M_{tot} = 8$ M$_\Earth$, $a = 0.30$ AU, $R_i = 10$ R$_\Earth$, and $Z/Z_\odot = 30$, and the specific models are those with initial $f_{env}$ = 0.01 and 0.05. Both cases show negligible change in the rock-iron core over time, while the convective envelope gradually shrinks due to contraction in both cases (while also losing some mass), as indicated by the increasing density near the core-envelope boundary. The temperature of the core and the temperature profile of the envelope also decrease significantly over time.

\section{Example Mass Loss Rates}
\label{mass_loss}

\begin{figure}[htp]
\includegraphics[width=\columnwidth]{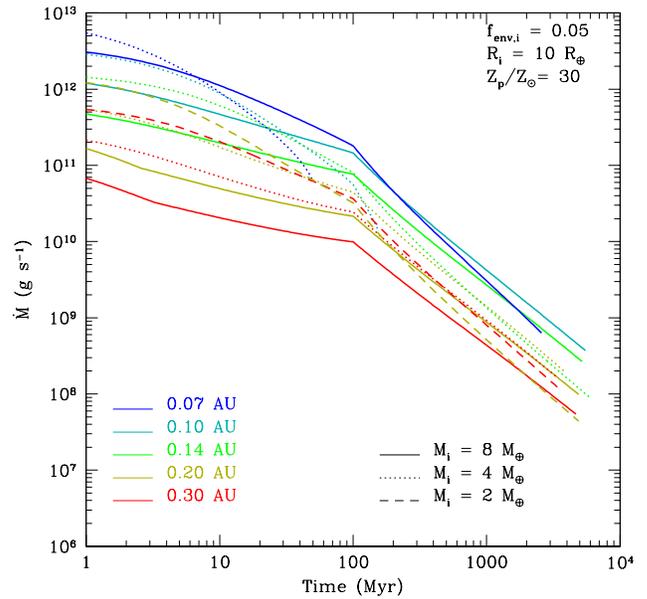}
\caption{Plot of the mass loss rates of mini-Neptune models with $a=0.30$ AU, and our ``default'' parameters of $R_i = 10$ R$_\Earth$, $Z/Z_\odot=30$, and $\epsilon = 0.10$. Orbital distances of 0.07 AU, 0.10 AU, 0.14 AU, 0.20 AU, and 0.30 AU are plotted with an initial envelope fraction of $f_{env} = 0.05$ and initial total masses of 2, 4, and 8 M$_\Earth$.}
\label{mdot-dist}
\end{figure}

In Figure \ref{mdot-dist}, we provide mass loss rates over time for some sample planet models. We compute models with initial total masses of 2 $M_\Earth$, 4 $M_\Earth$, and 8 $M_\Earth$, and orbital distances of 0.07 AU, 0.10 AU, 0.14 AU, 0.20 AU, and 0.30 AU, with an initial envelope fraction of $f_{env} = 0.05$ and with initial radii of 10 R$_\Earth$. Given the relation $\dot{M}\propto\frac{F_{XUV}R^3}{K_{tide}}$, we expect the mass loss rates to fall off roughly in line with the XUV flux, with smaller influences from the decreasing radii and tidal corrections. This is indeed what we see, with the mass loss rates being flat or varying relatively little at $t<100$ Myr, the time during which XUV flux is constant, decreasing for models that undergo significant contraction during this period. At $t>100$ Myr, the mass loss rates fall off at on the order of the same rate as the XUV flux. We also see mass loss rates decreasing with increasing orbital distances due to the lower XUV flux.

\begin{figure}[htp]
\includegraphics[width=\columnwidth]{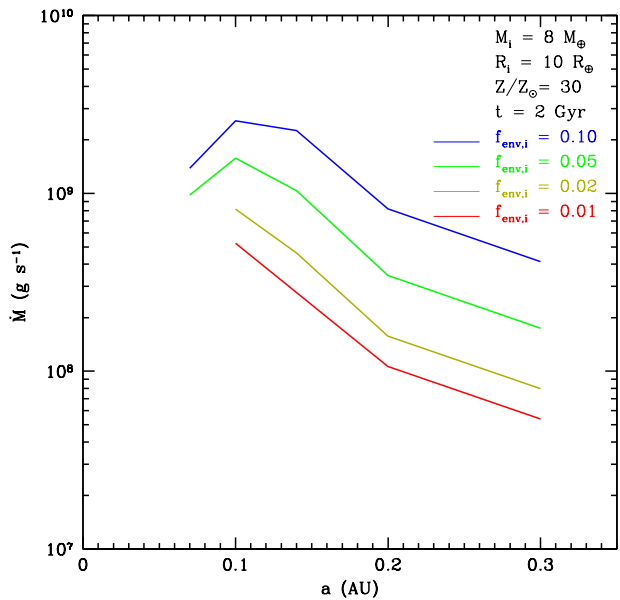}
\caption{Plot of mass loss rate versus orbital distance at an age of 2 Gyr for mini-Neptune models. We include models with $R_i = 10$ R$_\Earth$, $M_i = 8$ M$_\Earth$, and $f_{env,i} = $ 0.01, 0.02, 0.05, and 0.10.}
\label{mdotvsdist}
\end{figure}

As a further illustration, in Figure \ref{mdotvsdist} we plot the mass loss rate versus distance at 2 Gyr for select models, where we set $M_i = 8$ M$_\Earth$, and we plot $f_{env,i}$ of 0.01, 0.02, 0.05, and 0.10. Here the behavior is as expected with $\dot{M}\propto a^{-2}/K_{tide}$ from about 0.10 AU to 0.30 AU. The turnoff at the closest distances, 0.07-0.10 AU, occurs because most of the envelope has been lost at this point, decreasing the radius of the planet and, thus, the mass loss rate.

\section{Parameter Study}
\label{parameter_study}

\begin{figure*}[htp]
\includegraphics[width=\textwidth]{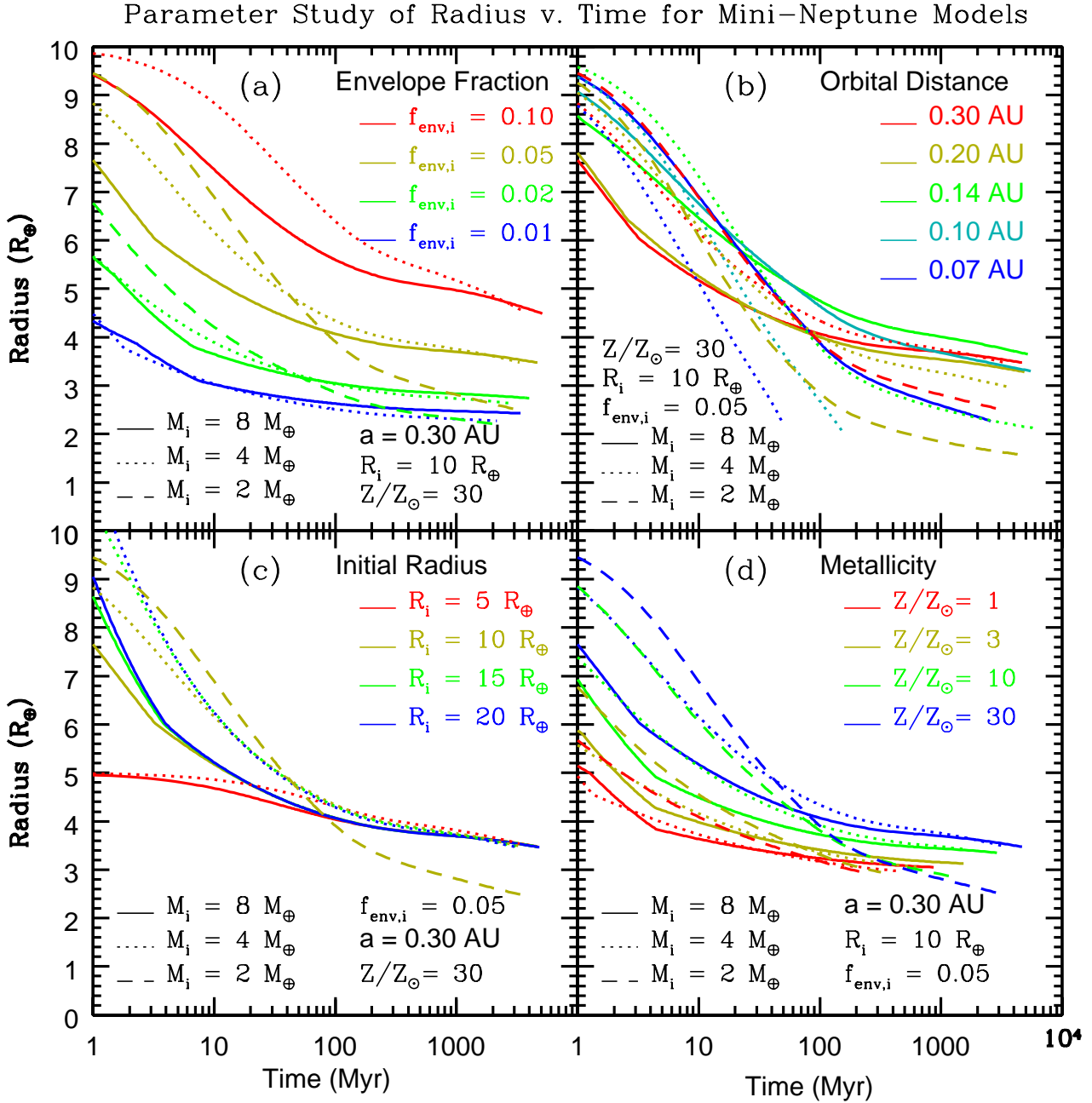}
\caption{Parameter study of radius evolution of mini-Neptune models. In all cases, models with initial masses of 2, 4, and 8 M$_\Earth$ are shown. Panel (a): study of initial envelope fraction with $f_{env,i} = $ 0.01, 0.02, 0.05, and 0.10. Panel (b): study of orbital distance with $a = $ 0.07, 0.10, 0.14, 0.20, and 0.30 AU. Panel (c): study of initial radius with $R_i = $ 5, 10, 15, and 20 R$_\Earth$. Panel (d): study of metallicity with $Z/Z_\odot = $ 1, 3, 10, and 30.}
\label{rad-params}
\end{figure*}

\begin{figure*}[htp]
\includegraphics[width=\textwidth]{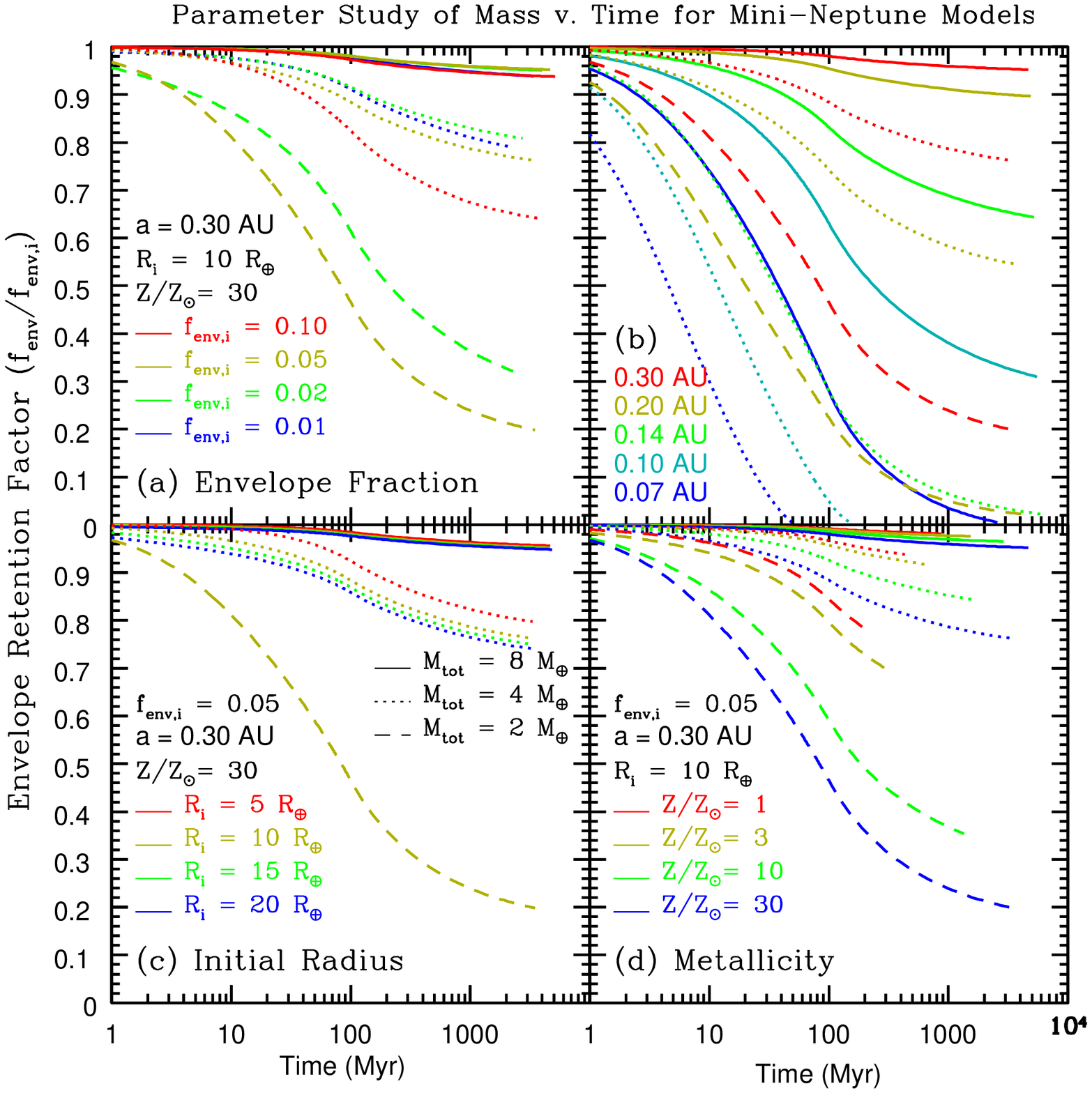}
\caption{Parameter study of envelope evaporation of mini-Neptune models. In all cases, models with initial masses of 2, 4, and 8 M$_\Earth$ are shown. Panel (a): study of initial envelope fraction with $f_{env,i} = $ 0.01, 0.02, 0.05, and 0.10. Panel (b): study of orbital distance with $a = $ 0.07, 0.10, 0.14, 0.20, and 0.30 AU. Panel (c): study of initial radius with $R_i = $ 5, 10, 15, and 20 R$_\Earth$. Panel (d): study of metallicity with $Z/Z_\odot = $ 1, 3, 10, and 30.}
\label{mass-params}
\end{figure*}

\begin{figure*}[htp]
\includegraphics[width=\textwidth]{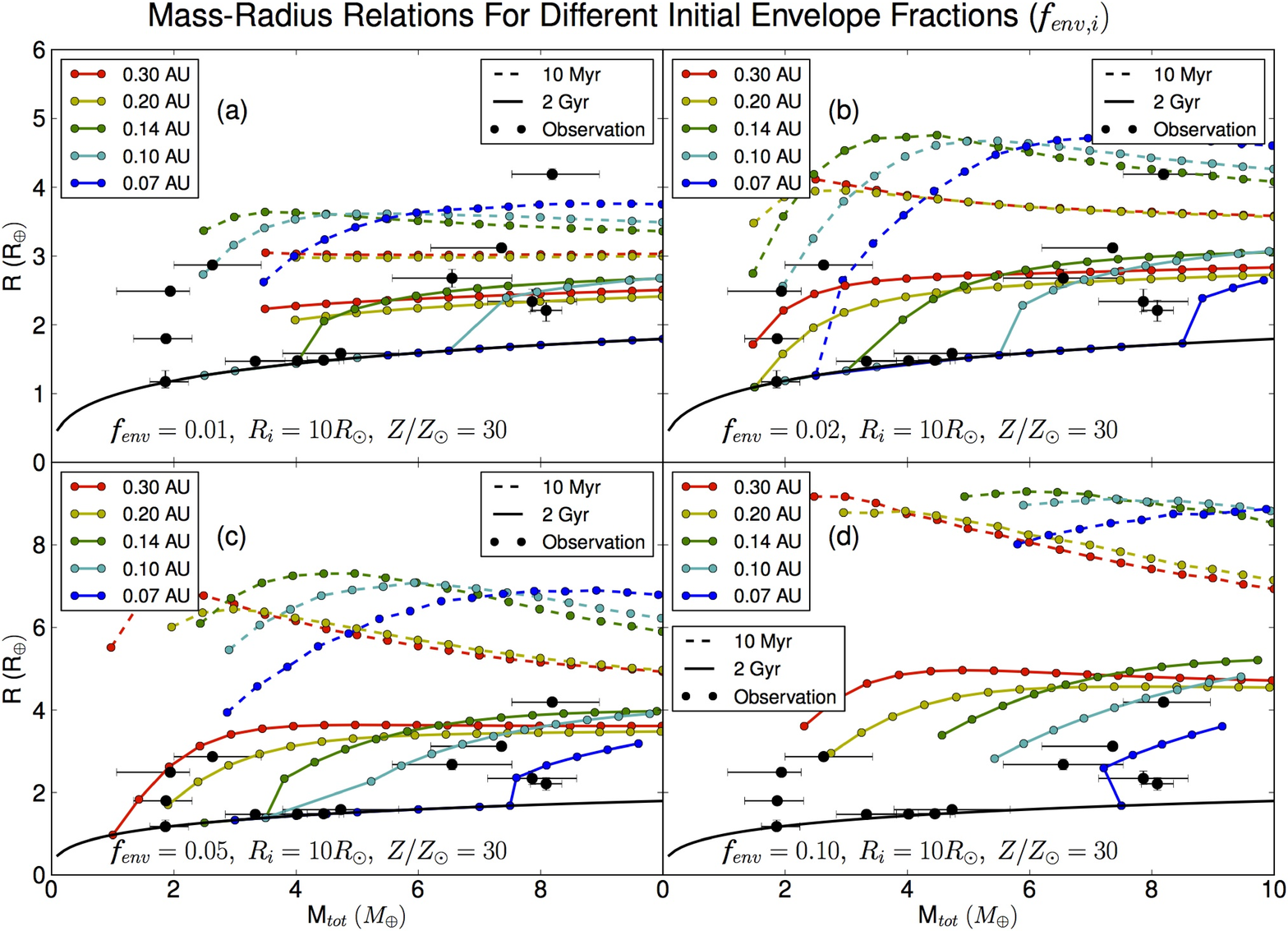}
\caption{Study of mass-radius curves with different $f_{env,i}$ at 10 Myr and 2 Gyr for mini-Neptune models. In all cases, models with $a = 0.30$ AU, and $R_i = 10$ R$_\Earth$, $Z/Z_\odot = 30$, and initial total masses of 0.5-10 M$_\Earth$ are plotted. Panel (a): $f_{env,i} = 0.01$. Panel (b): $f_{env,i} = 0.02$. Panel (c): $f_{env,i} = 0.05$. Panel (d): $f_{env,i} = 0.10$.}
\label{fenv}
\end{figure*}

\begin{figure}[htp]
\includegraphics[width=\columnwidth]{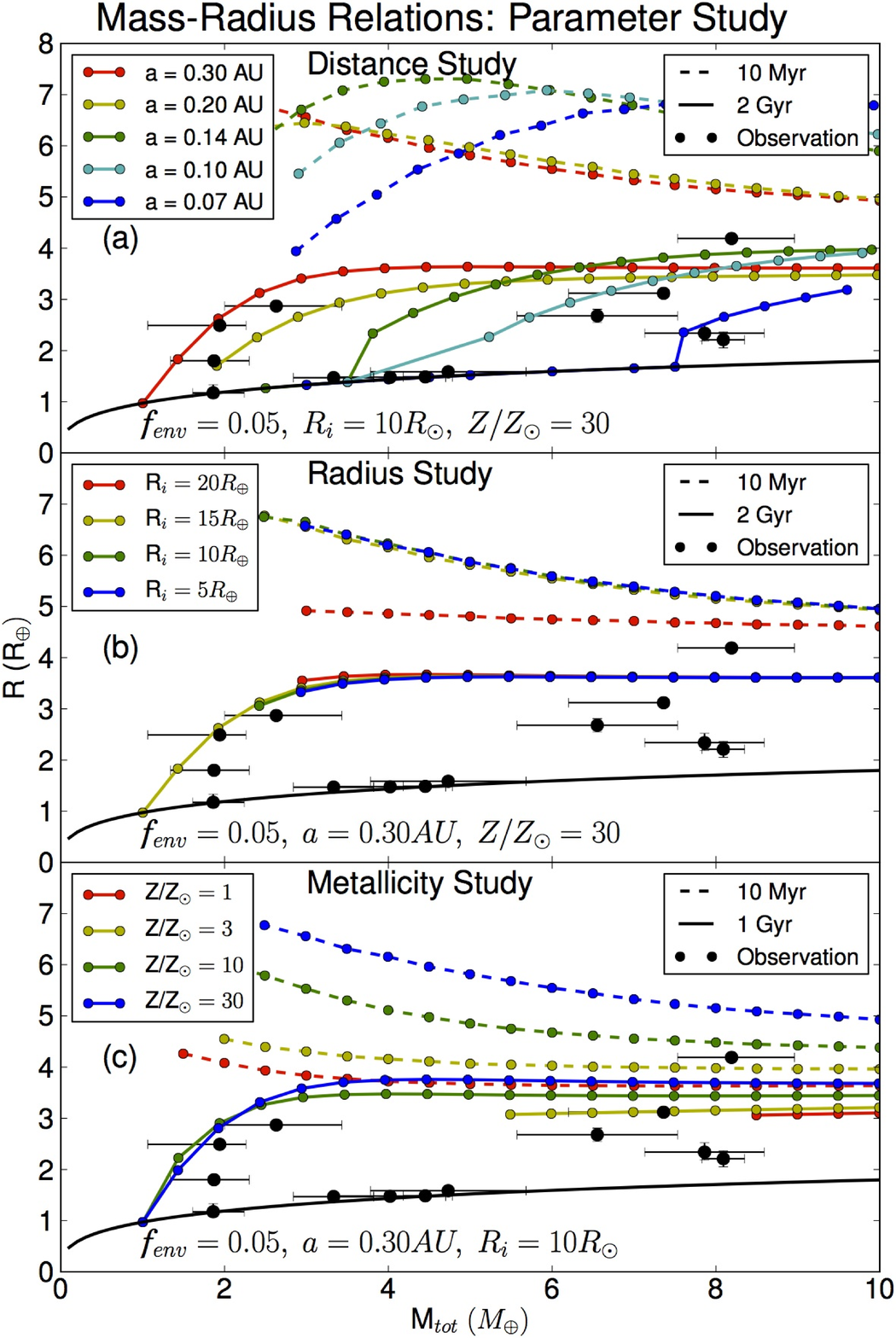}
\caption{Parameter study of mass-radius curves at 10 Myr and 2 Gyr for mini-Neptune models. In all cases, models with $f_{env,i} = 0.05$ and initial total masses of 0.5-10 M$_\Earth$ are plotted. Panel (a): study of orbital distance with $a = $ 0.07, 0.10, 0.14, 0.20, and 0.30 AU. Panel (b): study of initial radius with $R_i = $ 5, 10, 15, and 20 R$_\Earth$. Panel (c): study of metallicity with $Z/Z_\odot = $ 1, 3, 10, and 30.}
\label{params}
\end{figure}

In this section, we study in depth the effects of six parameters on the models produced by our code: initial total mass ($M_i$), initial envelope fraction ($f_{env,i}$), orbital distance ($a$), initial radius ($R_i$), metallicity ($Z/Z_\odot$), and mass loss prescription ($\epsilon$ and $t_M$). Figures \ref{rad-params}-\ref{mass-rad-ml} show the results of our calculations.

\subsection{Dependence on Initial Mass}

The mass of the solid (terrestrial-composition) core comprises most of the total mass in our models. In Figures \ref{rad-params} and \ref{mass-params}, we plot models with $M_i$ = 2, 4, and 8 M$_\Earth$. In Figures \ref{fenv}, \ref{params}, \ref{big}, and \ref{mass_loss}, we plot theoretical mass-radius relations for our models at different times over an initial mass range of 0.5-10 M$_\Earth$. We plot these models with a mass-radius relation for a purely terrestrial composition (black line) and observational data (black points) from \citet{2015ApJ...800..135D} and \citet{2015Weiss..in.press}. Note that our theoretical mass-radius curves are at fixed orbital distances, which are not the same as those of the observed planets. The observed planets have different ages and orbital distances than our model curves and serve simply to illustrate the range of planets we can model by comparison.

For comparison, low-density exoplanets have been observed with masses as small as $\sim$2 M$_\Earth$, while the largest terrestrial planets are $\sim$8 $M_\Earth$ (see Fig. \ref{mass-rad}), although this limit may be significantly greater for icy planets. Also, it is generally thought that 10 $M_\Earth$ is the critical core mass for runaway gas accretion, although there is some uncertainty in this, as suggested by \citet{2006ApJ...648..666R}. More recently, \citet{2014ApJ...797...95L}, found that 10-M$_\Earth$ rocky cores readily undergo runaway gas accretion regardless of orbital distance, but they speculate as to mechanisms that might prevent this and lead to the observed preponderance of super-Earths. Therefore, 2-10 M$_\Earth$ may be considered a plausible range of core masses.

We find that planets with lower initial masses experience more mass loss due to lower gravity, although the amount of mass loss is more sensitive to orbital distance  (see Section \ref{distance}). The lowest-mass planets lose their envelopes completely, although the limit of this effect also varies with orbital distance. However, for higher masses, the mass-radius curves are remarkably insensitive to initial mass. We also find that most of our models have ``final'' radii of 2-4 R$_\Earth$, the range where the greatest number of Kepler planet candidates have been observed.

\subsection{Dependence on Envelope Mass Fraction}

In Panel (a) of Figures \ref{rad-params} and \ref{mass-params} and all of Figure \ref{fenv}, we see the effect of varying the initial envelope mass fraction, $f_{env,i}$ on our model evolution, with $f_{env,i} = $ 0.01, 0.02, 0.05, and 0.10. Notably, for these and all of our models, we find that the radius evolution nearly ceases at Gyr ages because both cooling and mass loss are much slower then than at earlier times. This removes one important degeneracy when trying to fit theory to observed planets: for most observed planets, which are usually older than 1 Gyr, their properties will have very little dependence on their age.

Meanwhile the ``final radii'' (or radii at Gyr ages) can vary significantly depending on the masses of both the solid core (which constitutes most of the total mass) and the gaseous component at formation. For the most part, this applies to $f_{env,i}$ in particular; the mass-radius curves are remarkably flat above a certain mass, with variations in radius of $\lesssim 0.1$ R$_\Earth$ over a wide range of masses. In any case, the difference in radius at Gyr ages can be multiple Earth radii with different initial envelope fractions.

We also find that the amount of mass loss as a fraction of the initial envelope mass varies relatively little with $f_{env}$, especially at higher masses (e.g., $\sim$8 M$_\Earth$), where it is effectively negligible. However, this factor is more important at lower masses.

\subsection{Dependence on Orbital Distance}
\label{distance}

In Panel (b) of Figures \ref{rad-params} and \ref{mass-params}, and Panel (a) of Figures \ref{params}, we present models with a range of orbital distances: 0.07, 0.10, 0.14, 0.20, and 0.30 AU. This is the distance range (stellar flux range) in which most of the mini-Neptunes with significant envelopes have been discovered. From these, we find that the {\it shape} of our theoretical mass-radius curves (at least at Gyr ages) is similar across our models, but shifted left or right, to lower or higher masses, depending upon orbital distance. This is most readily seen in Panel (a) of Figure \ref{params}.

For a fixed $f_{env,i}$, the mass-radius curve at 2 Gyr is approximately flat at high masses, while at lower masses, it reaches a ``turnoff point'' that depends on the orbital distance, at which the radius rapidly decreases over a range of roughly 2 M$_\Earth$ until it reaches the terrestrial core (complete envelope evaporation). For closer orbital distances, this turnoff point occurs at higher masses, approximately 1-2 M$_\Earth$ higher for each doubling in stellar irradiation. (It also occurs at somewhat lower masses for higher $f_{env,i}$.)

The radius at 2 Gyr is remarkably consistent over a range of masses, and, therefore, over a range of envelope masses and surface gravities as well. To investigate how this relates to the amount of mass loss, in Panel (b) of Figure \ref{mass-params}, we plot the envelope mass evolution, that is the fraction of the mass of the envelope remaining, versus time for selected models (all with $f_{env,i} = 0.05$) at a range of orbital distances. Many of the resultant curves represent models that converge to approximately the same radius at Gyr ages, but nonetheless undergo very different amounts of mass loss, losing anywhere from 5\% to nearly 100\% of their envelopes. Thus, the collapse of the envelope due to cooling is significantly more dependent on orbital distance than on other parameters, regardless of the density and pressure in the envelope. The pattern of mass loss increasing with decreasing orbital distance is as expected from our model, where the mass loss rate scales as $a^{-2}/K_{tide}$ (where $K_{tide}$ is the tidal correction).

\begin{figure}[htp]
\includegraphics[width=\columnwidth]{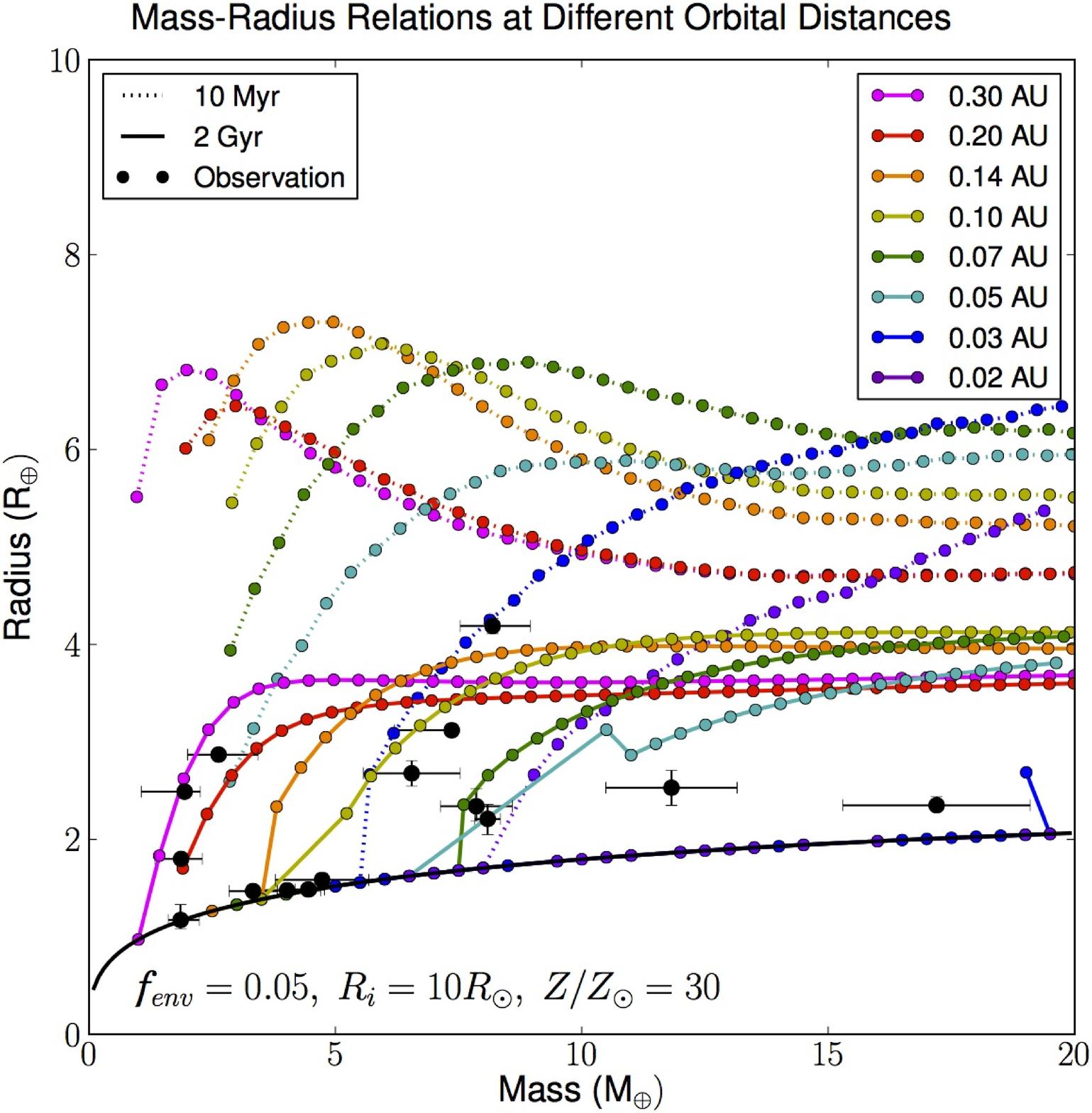} 
\caption{Mass-radius curves for planetary models with large masses of 10-50 M$_\Earth$, initial $f_{env}$ = 0.01, 0.02, 0.05, 0.10, and 0.20, initial radii of 10 R$_\Earth$, and orbital distances of 0.30 AU, plotted with observational data from \citet{2015ApJ...800..135D} and \citet{2015Weiss..in.press}. Curves at 100 Myr and 5 Gyr are shown.}
\label{big}
\end{figure}

In Figure \ref{big} we extend our plot of mass-radius curves to 20 M$_\Earth$ and orbital distances of 0.02 AU to 0.30 AU. Here, we find that the above trend in the shape of the mass-radius curve extends to both of these parameters, to some extent. On the high-mass end, the mass-radius curves remain approximately flat to 20 M$_\Earth$. Meanwhile, the previously-produced behavior of a turnoff point (where, for lower masses, the radius decreases rapidly) continues to an orbital distance of 0.05 AU, where it occurs at a higher mass and leaves higher-mass bare cores. However, for 0.03 AU, complete evaporation of the envelope occurs almost all the way up to 20 M$_\Earth$. This is consistent with the fact that no planets with likely gaseous envelopes are observed with stellar fluxes greater than $\sim$700 F$_\Earth$, equivalent to about $a = 0.04$ AU around a sun-like star.

\begin{figure}[htp]
\includegraphics[width=\columnwidth]{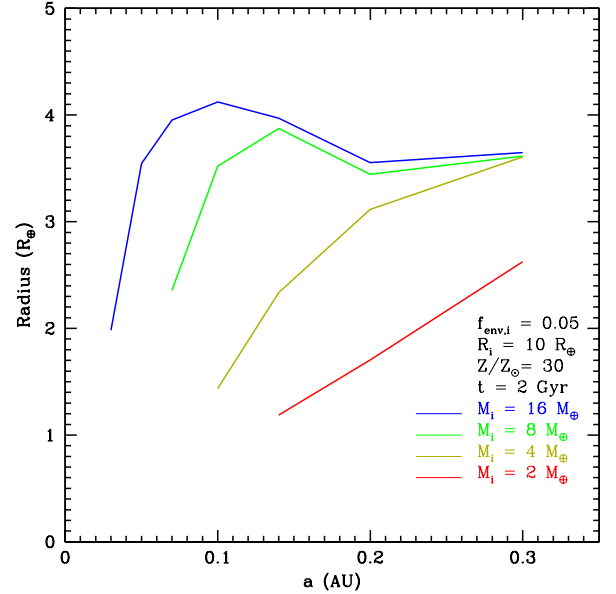}
\caption{Plot of radius versus orbital distance at an age of 2 Gyr for mini-Neptune models. We include models with $R_i = 10$ R$_\Earth$, $f_{env,i} = 0.05$ M$_\Earth$, and $M_i = $ 2, 4, 8, and 16 M$_\Earth$.}
\label{rvsdist}
\end{figure}

\begin{figure}[htp]
\includegraphics[width=\columnwidth]{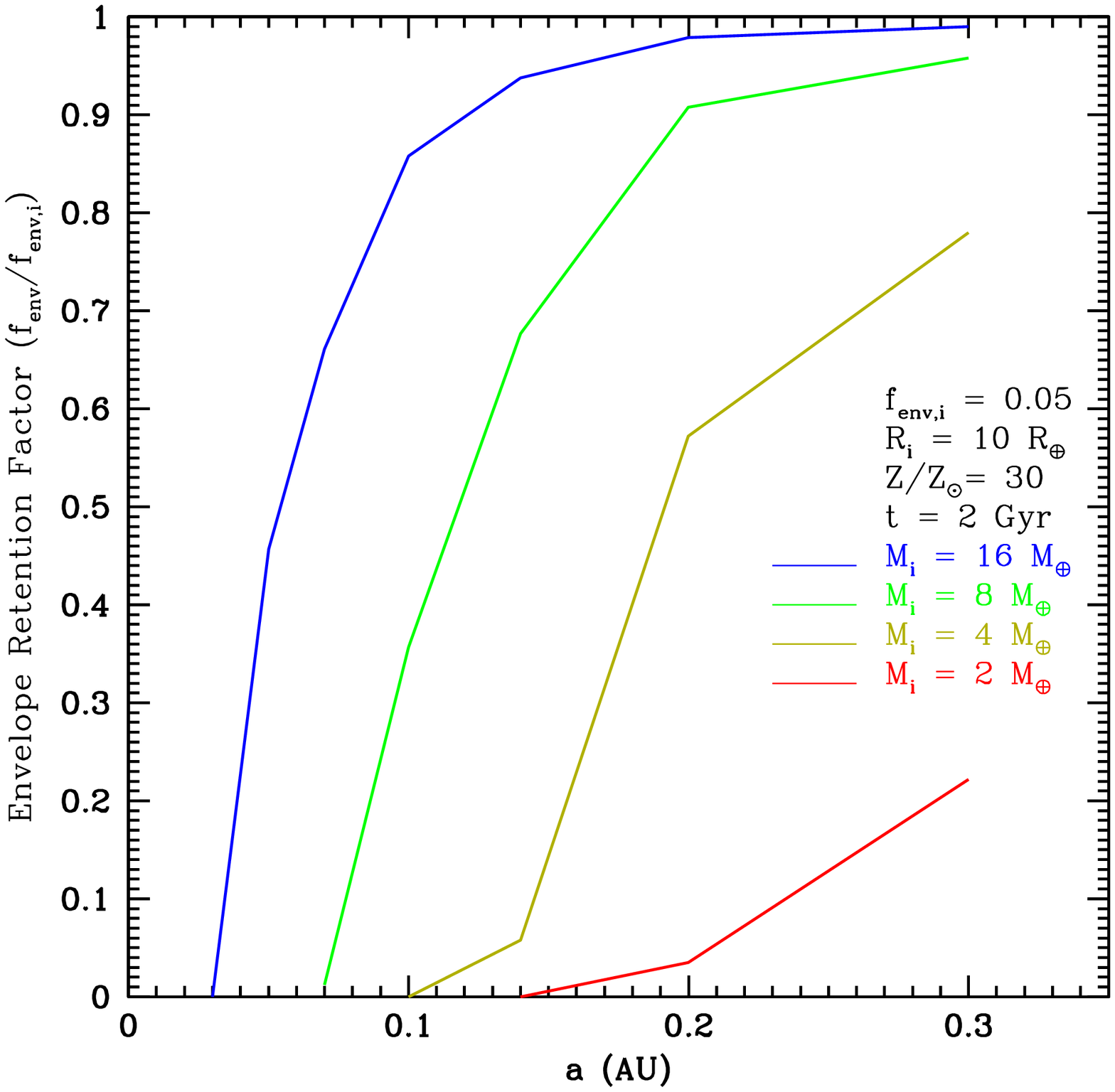}
\caption{Plot of envelope retention factor versus orbital distance at an age of 2 Gyr for mini-Neptune models. We include models with $R_i = 10$ R$_\Earth$, $f_{env,i} = 0.05$, and $M_i = $ 2, 4, 8, and 16 M$_\Earth$.}
\label{mlvsdist}
\end{figure}

To further illustrate the effects of orbital distance, in Figures \ref{rvsdist} and \ref{mlvsdist}, we plot several quantities for representative mini-Neptune models versus orbital distance over a range of 0.07-0.30 AU. In each case, we plot the states of the models at an age of 2 Gyr, and we plot models with $f_{env,i} = 0.05$ and $M_i = $ 2, 4, 8, and 16 M$_\Earth$, otherwise using our default parameters.

Figure \ref{rvsdist} shows a plot of radius at 2 Gyr versus orbital distance. We again see that the radius is relatively flat with orbital distance at lower levels of irradiation, but rapidly falls off at closer orbital distances because the envelope has mostly or entirely evaporated. This turnoff occurs at 0.10 AU for the 16-M$_\Earth$ models, but farther out for lower masses to the point where it occurs outward of 0.30 AU for the 2-M$_\Earth$ models.

Figure \ref{mlvsdist} plots the fraction of the envelope mass remaining after 2 Gyr versus orbital distance. This quantity follows a general pattern of increasing from zero or near zero to a high level over a small factor of 2-3 in orbital distance. For example, the envelope retention factor for the 16-M$_\Earth$ models increases from zero at 0.03 AU to $\sim$0.85 at 0.10 AU. At the same time, this same shift occurs much farther out for the 2-M$_\Earth$ models, with an envelope retention factor of $\sim$0.03 at 0.20 AU and $\sim$0.22 at 0.30 AU, retaining a majority of the initial envelope only at larger distances.

\subsection{Dependence on Initial Radius}
\label{init_rad}

The initial radius of a given planet at the end of accretion is unclear. Therefore, we explore models of planets with H$_2$-He envelopes with initial radii in the range from 22.6 R$_\Earth$\footnote{Equivalent to 2 Jupiter radii.}, near the probable maximum size of exoplanets of any mass \citep{2012A&A...547A.112M,2010ApJ...709..159A}, down to a radius of $\sim$2-3 R$_\Earth$.\footnote{The actual limit is $T_{eff}=50$ K, the boundary of our atmosphere grid, which corresponds to a radius of 2-3 R$_\Earth$ for some of our representative models.}

In Panel (c) of Figures \ref{rad-params} and \ref{mass-params}, and Panel (b) of Figures \ref{params}, we present models with a range of initial radii: 5, 10, 15, and 20 R$_\Earth$. $R_i = 10$ R$_\Earth$ produces the widest range of well-defined models, which is why we have used it as a default throughout our parameter study.

By the assumptions of our model, mass loss is proportional to the amount of starlight intercepted by the planet and is inversely proportional to the gravitational potential at the XUV photosphere, thus making it proportional to $R^3/K_{tide}$. Therefore a larger initial radius leads to much greater mass loss and could even lead to complete evaporation of the envelope. ($1/K_{tide}$ also increases with radius.)

Notably, however, we find that the initial radius of the models makes only a small difference in either radius or mass loss at Gyr ages. This is because of rapid radius contraction at early times, causing models differing in initial radii to converge or nearly converge by an age of 100 Myr, and often much earlier. This early convergence in radius also results in very small differences in mass loss, which is primarily dependent on radius for equal-mass planets. The exceptions to this rule are models with low initial total mass, which fall below the turnoff point. This removes one source of degeneracy in trying to fit observational data.

\subsection{Dependence on Metallicity}

In Panel (d) of Figures \ref{rad-params} and \ref{mass-params}, and Panel (c) of Figures \ref{params}, we present models with a range of envelope metallicities: 1x, 3x, 10x, and 30x solar metallicity. For comparison, atmospheres of super-Earths and mini-Neptunes are expected to be significantly enriched compared with solar. Specifically, Uranus and Neptune are enriched to $\sim$50 times solar values \citep{2011ApJ...726...15H}. While it is not obvious {\it a priori} that close-in planets should be similarly enriched, it is a reasonable assumption since observations of transiting super-Earths and sub-Neptunes support either high-altitude clouds and/or a high mean molecular weight atmosphere \citep{2012ApJ...756..176H}.

Varying the metallicity of the models has some effect, though not as large as that of initial total mass or orbital distance. A higher metallicity increases the mean molecular weight of the atmosphere, which should, in principle, increase the density. However, we find in our simulations that this is a small effect on the convective region and that the greater opacity, which results in a thicker radiative layer and slower contraction time, results in a larger overall radius at a given age for higher metallicity (despite higher mass loss). Thus, a higher metallicity results in a larger radius at Gyr ages. The difference in final radius between 1x and 30x solar metallicity is about 0.40 R$_\Earth$, although this effect is more dramatic at early times.

Because of this smaller variation, and because the atmospheres of mini-Neptune-type planets are expected to be highly enriched in general ($\gtrsim$30x solar), metallicity is not a critical parameter to fit observational data.

The effect of metallicity on mass loss is similar to the effect of initial envelope fraction. It is relatively unimportant compared with other parameters, and it is especially negligible for the higher mass objects in our models with $M_i = 8$ M$_\Earth$.

\subsection{Dependence on Mass Loss Prescription}

\begin{figure}[htp]
\includegraphics[width=\columnwidth]{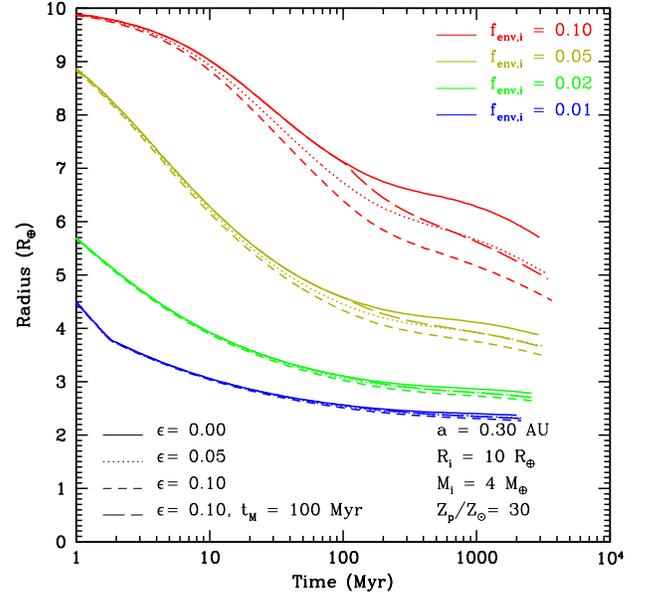}
\caption{Comparison of different mass loss prescriptions in plots of radius versus time for mini-Neptune models. Plotted are $\epsilon = 0$ (no mass loss), $\epsilon = 0.05$, $\epsilon = 0.10$, and $\epsilon = 0.10$ with the mass loss switched off until 100 Myr. We include models with $R_i = 10$ R$_\Earth$, $a = 0.30$ AU, $M_i = 4$ M$_\Earth$, and $f_{env,i} = $ 0.01, 0.02, 0.05, and 0.10.}
\label{rad-ml}
\end{figure}

\begin{figure}[htp]
\includegraphics[width=\columnwidth]{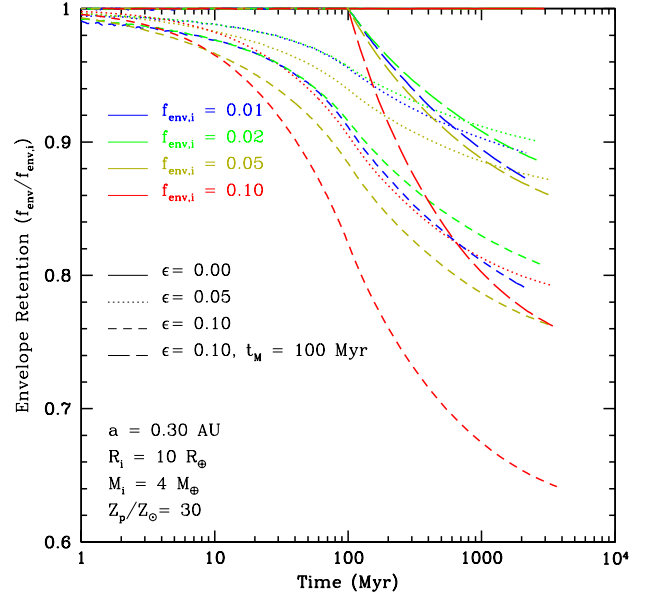}
\caption{Comparison of different mass loss prescriptions in terms of the mass fraction of the original envelope remaining versus time for mini-Neptune models. Plotted are $\epsilon = 0$ (no mass loss), $\epsilon = 0.05$, $\epsilon = 0.10$, and $\epsilon = 0.10$ with the mass loss switched off until 100 Myr. We include models with $R_i = 10$ R$_\Earth$, $a = 0.30$ AU, $M_i = 4$ M$_\Earth$, and $f_{env,i} = $ 0.01, 0.02, 0.05, and 0.10.}
\label{mass-ml}
\end{figure}

While the relationships between mass loss rates and physical parameters are straightforward, the processes that lead to mass loss in gaseous exoplanets are uncertain \citep{2009ApJ...693...23M}. However, especially for a specific model, the mass loss behavior may be parameterized by the efficiency$-$the fraction of intercepted XUV energy that goes into mass loss. In most of our models we set the efficiency at $\epsilon = 0.10$. We also have the option of having mass loss initially switched off and then turned on at some point in time, $t_M$. This simulates a planet that forms at a relatively large distance, where mass loss is negligible, and then migrates to a close-in orbit where it undergoes higher mass loss at time $t_M$.

In Figures \ref{rad-ml} and \ref{mass-ml}, we plot radius and mass, respectively, versus time for planet models with four different mass loss prescriptions: our default $\epsilon = 0.10$, $\epsilon = 0.05$, $\epsilon = 0$ (no mass loss), and a fourth model with $\epsilon = 0.10$ switched on at $t_M=100$ Myr. This type of prescription can bracket the actual, smoother variation in mass loss rates for migrating planets. We plot models with $f_{env}$ = 0.01, 0.02, and 0.05 and $a = 0.30$ AU, and we otherwise use our default parameters.

With no mass loss, the only influence on the radius is cooling, and the final radii are significantly larger than with mass loss, by up to about an Earth radius. Notably, without mass loss, the envelopes of low-mass planets never shrink to small radii, and so they are not consistent with low-mass, low-density planets such as Kepler-11b, -11c, and -11f. Some mass loss is required to fit our models to observations in this part of the parameter space.

Setting $\epsilon = 0.05$ leads to half as much mass loss as our default models and makes up approximately half the difference in radius. The models in which the mass loss switches on only after 100 Myr also experience about half as much mass loss, both in terms of radius and actual mass. Late migration due to planet-planet scattering or the Kozai mechanism beginning at $t\gtrsim 200$ Myr would have a much greater impact on mass loss because the XUV radition falls off as $t^{-1.23}$ after 100 Myr.

\begin{figure*}[htp]
\includegraphics[width=\textwidth]{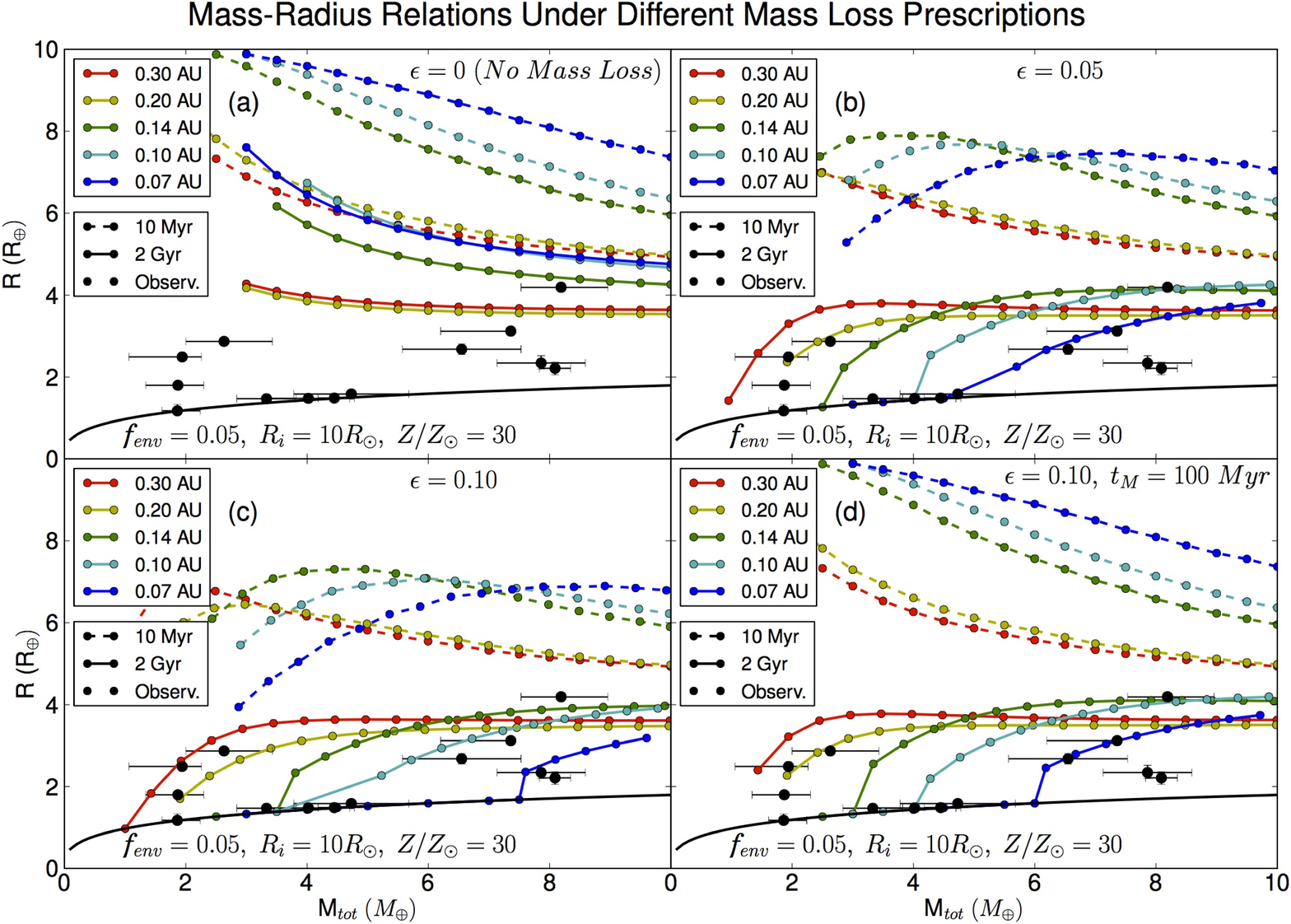}
\caption{Mass-radius curves at 10 Myr and 2 Gyr for mini-Neptune models with different mass loss prescriptions, plotted with a terrestrial, Earth-like composition model (black line), and observational data (black points) from \citet{2015ApJ...800..135D} and \citet{2015Weiss..in.press}. In all cases, models with $R_i = 10$ R$_\Earth$, $f_{env,i} = 0.05$, orbital distances of 0.07, 0.10, 0.14, 0.20, and 0.30 AU, and initial masses of 0.5-10 M$_\Earth$ are plotted. Panel (a): $\epsilon = 0$ (no mass loss). Panel (b): $\epsilon = 0.05$. Panel (c): $\epsilon = 0.10$. Panel (d): $\epsilon = 0.10$ with the mass loss switched off until 100 Myr.}
\label{mass-rad-ml}
\end{figure*}

In Figure \ref{mass-rad-ml}, we plot mass-radius curves over a range of 0.5-10 M$_\Earth$ for the same four cases. For this figure, we plot the mass-radius curves at 10 Myr and 2 Gyr. We also plot a range of orbital distances: 0.07, 0.10, 0.14, 0.20, and 0.30 AU (colored lines). We find that for all of our mass loss prescriptions except the case of no mass loss, several of the 2 Gyr curves pass through the low-density (low mass and high radius) planets on the plots, but they do not necessarily correspond to the same orbital distances. We investigate fitting our models to specific observed planets in Section \ref{kepler-11}.


\section{Application to the Kepler-11 System}
\label{kepler-11}

\begin{table*}[tbhp]
\caption{Mass and Radius Observations of the Kepler-11 Planets}
\begin{center}
\begin{tabular}{l|l|l|l|l}
\hline
Planet                   & Orbital Distance (AU) & Radius (R$_\Earth$) from    & Our adopted mass (M$_\Earth$) from & Mass (M$_\Earth$) adopted by \\
                         &                       & \citet{2013ApJ...770..131L} & \citet{2015Weiss..in.press}           & \citet{2012ApJ...761...59L}  \\
\hline
Kepler-11b               & 0.091 & $1.80\pm 0.04$              & $1.87_{-0.53}^{+0.43}$            &  $4.3_{-2.0}^{+2.2}$         \\
Kepler-11c               & 0.107 & $2.87\pm 0.05$              & $2.63_{-0.63}^{+0.80}$            & $13.5_{-6.1}^{+4.8}$         \\
Kepler-11d               & 0.155 & $3.12\pm 0.06$              & $7.36_{-1.16}^{+0.03}$            &  $6.1_{-1.7}^{+3.1}$         \\
Kepler-11e               & 0.195 & $4.19\pm 0.08$              & $8.19_{-0.66}^{+0.77}$            &  $8.4_{-1.9}^{+2.5}$         \\
Kepler-11f               & 0.250 & $2.49\pm 0.05$              & $1.94_{-0.88}^{+0.32}$            &  $2.3_{-1.2}^{+2.2}$         \\
Kepler-11g               & 0.466 & $3.33\pm 0.07$              & $5.20^{+27.6}$                    &  N/A                         \\
\hline
\end{tabular}
\end{center}
\label{kepler11data}
\end{table*}

Kepler-11 is perhaps the best- studied planetary system in terms of the structures of its planets. It is a densely-packed system with six known planets with semi-major axes ranging from 0.091 AU to 0.466 AU \citep{2011Natur.470...53L}. The masses of five of these planets have been precisely measured with transit timing variations \citep{2013ApJ...770..131L} and, more recently, with radial velocities \citep{2015Weiss..in.press}. The measured orbital distances, masses, and radii for these planets are shown in Table \ref{kepler11data}.

\citet{2012ApJ...761...59L} studied the compositions of these planets using an evolutionary model with a H$_2$-He envelope surrounding an Earth-like core. However, these models were fit to the previous masses measured from TTVs, and the improved RV measurements invite a reanalysis of these planets. The measured masses for Kepler-11d, -11e, and -11f changed by a relatively small amount between the TTV and RV observations, but the masses (and densities) of Kepler-11b and -11c were revised down dramatically, by more than a factor of 2 (with significantly smaller error estimates), so their estimated compositions will be very different.

We can change the stellar type in our model, which we have done for the planets in the Kepler-11 system. Changing the central star in our model has two effects. First, the total thermal irradiation changes according to $a^{-2}R_*^2T_*^4$. Second, different stars will have different ultraviolet fluxes. \citet{2012ApJ...761...59L} assumed that all planet host stars have the same XUV flux evolution. However, other models suggest different prescriptions. \citet{2014AJ....148...64S} find that the power-law decay of XUV flux only holds at $\gtrsim 1$ Gyr for M dwarfs, and these stars begin with a lower XUV flux, which remains constant longer. However, Kepler-11 is a G dwarf, so this is not a concern.

A lower irradiation level would allow for more rapid cooling and contraction, leading to less mass loss, while a lower XUV flux also results in less mass loss. Therefore, around smaller stars, low-mass planets could retain gaseous envelopes longer and at closer distances than around larger stars.

We find in general that the most important parameters to fit the data are initial mass ($M_i$), initial envelope fraction ($f_{env,i}$), and orbital distance ($a$, which is known from observations), while we can choose $R_i = 10$ R$_\Earth$, $Z/Z_\odot = 30$, and $t\geq 2$ Gyr without loss of generality, although we use a larger initial radius for the lower mass planets, Kepler-11b, -11c, and -11f, which produces better convergence with our code.\footnote{Note that the evolutionary models used by \citet{2012ApJ...761...59L} employed $R_i = 10-15$ R$_\Earth$, except for Kepler-11b, for which $R_i\sim 30$ R$_\Earth$ was used.}

Kepler-11d and -11e are the easiest objects to interpret, having high enough masses and large enough orbital distances to retain significant amounts of hydrogen and helium over several Gyr. Given how closely the initial envelope fraction correlates with radius at Gyr ages in our model set, it is possible to determine $f_{env,i}$, or the initial envelope {\it mass}, $M_{env,i}$ fairly precisely. The uncertainty in the observed mass of 5\%-10\% introduces an uncertainty of similar size into $M_{env,i}$, but the fit of $M_{env,i}$ is positively correlated with the assumed mass of the model, so this uncertainty is diminished in $f_{env,i}$. Also, the roughly 2\% uncertainty in radius introduces a 5\%-10\% uncertainty into both $M_{env,i}$ and $f_{env,i}$.

For Kepler-11d, we find an initial H$_2$-He content of $0.25\pm 0.03$ M$_\Earth$, corresponding to $f_{env,i} = 3.7\pm 0.4\%$. Mass loss in this model is modest. At an age of 4 Gyr, which will be very similar to the result at the system's age of 8.5 Gyr, the resultant envelope fraction is $f_{env} = 2.6\pm 0.3\%$. This is in contrast with \citet{2012ApJ...761...59L}, who find a present-day envelope fraction of $f_{env} = 8.4_{-2.4}^{+2.7}\%$ for Kepler-11d (with older, but similar mass measurements).

For Kepler-11e, we find an initial H$_2$-He content of $0.68\pm 0.06$ M$_\Earth$, corresponding to $f_{env,i} = 8.3\pm 0.4\%$. At an age of 4 Gyr, which will be very similar to the result at the system's age of 8.5 Gyr, the resultant envelope fraction is $f_{env} = 6.5\pm 0.5\%$. \citet{2012ApJ...761...59L} find a present day $f_{env}$ for Kepler-11e of $17.2_{-4.2}^{+4.1}\%$ for Kepler-11e. For Kepler-11d and -11e, they estimate the initial envelope fraction, $f_{env,i}$ at about 28\%. Both of our estimates for present-day envelope fractions are lower by a factor of three.

Part of this discrepancy could conceivably result from our treatment of the radiative atmosphere. (Specifically, the thickness of the atmosphere, since the opacities affect the cooling rate and are a separate problem.) We determine an upper limit to the magnitude of this effect by fitting a model to Kepler-11d with no $\Delta R$ at all$-$that is, no radiative atmosphere. With this fit, we find an initial envelope fraction of $f_{env,i} = 5.4\pm 0.3\%$ and a present day envelope fraction of $f_{env,i} = 4.6\pm 0.2\%$, still a factor of two lower than the results of \citet{2012ApJ...761...59L}. Therefore, our treatment of the radiative atmosphere is likely not a large source of error in our analysis.

It is also possible that the difference in computed envelope fractions is partially due to the choice of equation of state, specifically, the use of the volume addition law for the envelope, and the density of the solid core (although the latter effect is likely small), both of which could affect the determination of the relative core-envelope fractions. However, the most important factor is likely the cooling rate of the models, based upon both the radiative properties of our atmospheric boundary conditions and the specific heat of the core, which contribute to the overall cooling at the same order of magnitude. Faster cooling results in more contraction and a larger envelope fraction to achieve the observed radii. Further work is needed to determine the magnitude of these effects and the most correct boundary condition.

The other three planets, Kepler-11b, -11c, and -11f, are much more difficult to interpret. Kepler-11b and -11c were both revised down in mass by more than a factor of two compared with the measurements used by \citet{2012ApJ...761...59L}, who already find a high mass loss rate, requiring a large initial envelope fraction. A lower mass, and thus a lower gravity, will increase this problem further.

In the case of Kepler-11b, the current radius of 1.80 R$_\Earth$ is too small to explain without evaporation of most, but not all, of its envelope. With such a low mass (1.87 M$_\Earth$) and a relatively high level of irradiation ($a = 0.091$ AU), the atmosphere puffs up enough that it is difficult to find a solution that achieves the observed radius. Furthermore, those solutions that do reach the observed radius are all close to evaporating completely by the present day, even if the evaporation efficiency is set an order of magnitude lower than expected, to $\epsilon = 0.01$, or if evaporation starts late due to migration ($t_M = 300$ Myr) due to planet-planet scattering or the Kozai mechanism. Thus, at expected evaporation rates, most models are likely to lose their envelopes completely, which is inconsistent with observations. Models with much higher starting masses (tens of M$_\Earth$) stay larger than the observed radius.

Because of its high irradiation ($a = 0.107$ AU), Kepler-11c is likely to be subject to the same problems as Kepler-11b in finding a solution that retains its envelope to the present time, although its mass is higher, which aids in envelope retention. Because its radius is larger, the ``water-world'' solution does not hold; its low density requires {\it some} hydrogen. For this object, we still find that all of our models lose their envelopes completely for our standard mass loss rate with an efficiency of $\epsilon = 0.10$. However, we find a solution of $f_{env,i}\sim 1.5\%$ when the mass loss efficiency is set to $\epsilon = 0.01$. For this solution, the present-day envelope mass fraction is $f_{env}\sim 0.25\%$, indicating a large amount of mass loss.

However, this seems to be an improbably low mass loss efficiency, so there may be a different mechanism at work. If we leave $\epsilon = 0.10$, but set $t_M = 300$ Myr, indicating that the planet moved to its current orbit from farther out at an age of 300 Myr, we find a similar solution with a somewhat larger envelope of $f_{env,i}\sim 0.025$ and a present-day envelope fraction of $f_{env}\sim 0.7\%$. We fail to produce such a result with a migration time significantly earlier than 300 Myr.

In light of this, we construct an alternative solution that explains the observations of both Kepler-11b and Kepler-11c. In this scenario, Kepler-11b is a ``water world'' that formed beyond the snow line with a composition consisting mostly of water, including an ice core, and either no hydrogen or helium or an envelope that has since evaporated completely, leaving a thick steam atmosphere. (This model is an estimate based on our previous work in \citealp{2012ApJ...756..176H} and \citealp{2014ApJ...787..173H}.) Kepler-11c also formed below the snow line as a mini-Neptune with an initial envelope mass fraction of $\sim$2.5\%. The two planets then moved to their current orbits some time after they formed, on the order of 300 Myr. In their current orbits, Kepler-11b lost any H$_2$-He envelope that it had, leaving its thick steam atmosphere, and Kepler-11c lost most of its envelope, reaching its current composition of $f_{env}\sim 0.7\%$. The water world explanation is the only model we find for Kepler-11b that fits the observations within observational uncertainty. However this fit is provisional, and additional research is needed to determine its plausibility.

For Kepler-11f, for the standard mass loss prescription of $\epsilon = 0.10$, we find an initial envelope fraction at the high-mass end of the uncertainty range of $f_{env,i} = 2.7\pm 0.2\%$. The models that best fit the observations fall near the edge of our grid of atmospheres, but we can extrapolate to a significantly higher fraction $f_{env,i}\sim 4\%$ near the middle of the uncertainty range. The present-day envelope fraction is on the order of $f_{env}\sim 0.25\%$, indicating significant mass loss. Both the initial and final envelope mass fractions in our analysis are an order of magnitude smaller than those found by \citet{2012ApJ...761...59L}.

The latest available data also include a mass measurement for Kepler-11g, which is 5.20 M$_\Earth$. The $2\sigma$ upper mass limit for this object is 27.6 M$_\Earth$. Assuming a true mass near this preliminary measurement, we find an initial envelope mass fraction for Kepler-11g of $f_{env,i} = 2.7\pm 0.3\%$. Because Kepler-11g is at a relatively large distance of 0.466 AU, the mass loss is small, with a present-day envelope fraction of $f_{env} = 2.6\pm 0.3\%$.


\section{Example Evolutionary Progressions of Transit Spectra}
\label{transit}

\begin{figure*}[htp]
\includegraphics[bb=18 400 592 718,width=\textwidth]{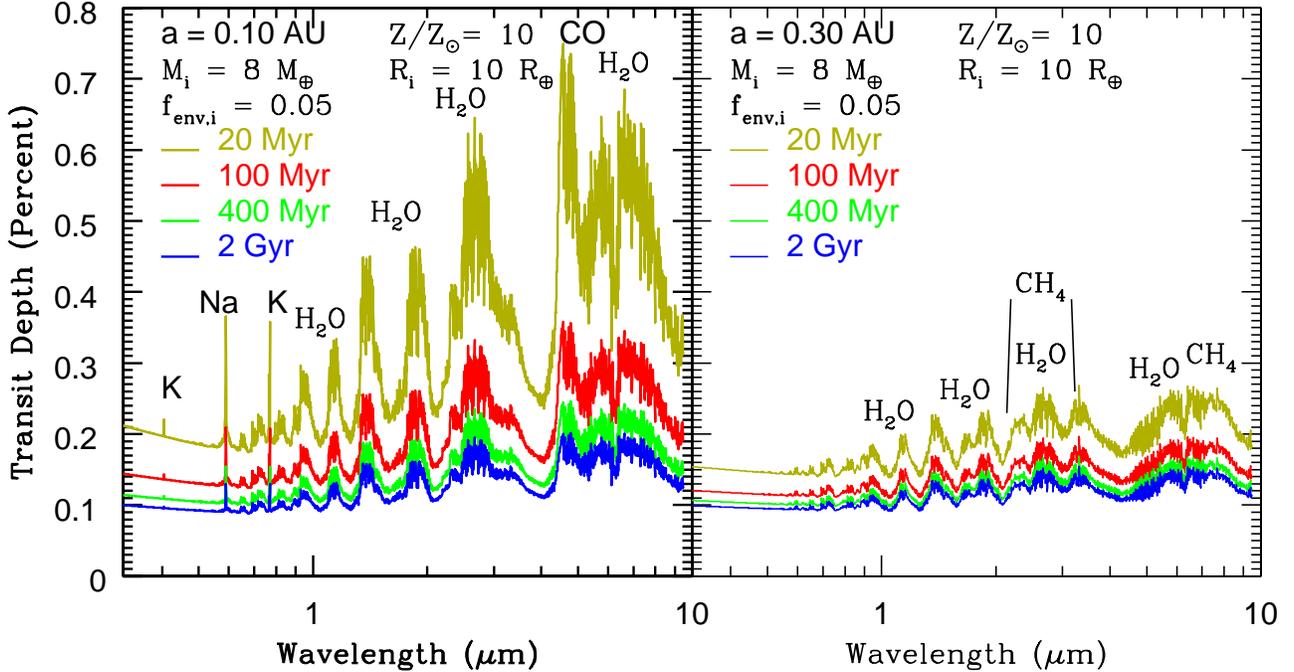}
\caption{Transit spectra over time for a planetary model with initial parameters of $M_{tot} = 8$ M$_\Earth$, $f_{env}$ = 0.05, and $R_i = 5$ R$_\Earth$, and $Z/Z_\odot = 10$. We assume a cloudless atmosphere (predicted by the simplest theoretical treatment of the given modeled atmosphere temperatures of 500-900 K for all models). Left panel: $a = 0.10$ AU; right panel: $a = 0.30$ AU.\vspace{18pt}}
\label{spectrum01}
\end{figure*}

Given the masses, radii, temperatures, and atmospheric compositions of our models over time, we can compute the evolution of their transit spectra using the methods of \citet{2012ApJ...756..176H}. In Figure \ref{spectrum01}, we present the evolution of two sample spectra. These models have parameters of initial $M_{tot} = 8$ M$_\Earth$, initial $f_{env} = 0.05$, $R_i = 10$ R$_\Earth$, and $Z/Z_\odot = 10$, and the specific models are those with $a$ = 0.10 (left panel) and 0.30 (right panel).

For the purpose of these spectral models, we consider an isothermal radiative atmosphere. We also model a cloudless atmosphere because that is most consistent with our evolutionary models. Observations of atmospheres of mini-Neptune-type planets remain ambiguous and may support a haze layer \citep{2012ApJ...756..176H}, as has been observed for giant planets. However, we note that our results were relatively little affected by metallicity, so the effect of clouds or hazes on evolution is likely to be similarly small. Moreover, this cloudless model is similar to that used in other evolutionary models such as those of \citet{2013ApJ...776....2L}.

The temperature of this isothermal model is equal to that of our full atmosphere model at a pressure level corresponding with a vertical Rosseland optical depth of $\tau = 0.56$, which is approximately the mean optical depth probed by the transit spectrum \citep{2008A&A...481L..83L}. Real atmospheric temperatures change little over a Gyr, primarily because of stellar irradiation, so the main sources of change in the spectra are the decrease in the overall transit depth due to contraction and the increase in gravity. As a result, the evolution of the spectra is mostly a flattening, and not a change, in the features.

However, between the two orbital distances, we see significant differences. First, the atmosphere temperatures provided by our code as described above are very different$-$nearly 800 K at 0.10 AU and only about 450 K at 0.30 AU. The higher temperature results in a higher scale height and a deeper atmosphere. Also, at the higher temperature, we see prominent Na and K alkali metal lines in the spectrum that we do not see at lower temperatures, and we see strong CO features and weak CH$_4$ features at 0.10 AU, whereas the opposite is true at 0.30 AU.


\section{Conclusions}
\label{conclusions}

We have constructed exoplanet evolution models for super-Earths and mini-Neptunes incorporating thermal cooling and XUV-driven mass loss over a wide range of parameters. With these models, we have conducted a parameter study of the effects of initial conditions: mass, radius, envelope mass fraction, orbital distance, and metallicity, as well as the particular prescription for mass loss on super-Earth and mini-Neptune evolution.

We find that the most important parameters to produce the observed variation in mini-Neptune type exoplanets are orbital distance, initial envelope mass fraction, and initial total mass, while age, metallicity, and initial radius make relatively little difference in the present-day observable properties of many such exoplanets, initial radius being the least important. Our general results show that a large subset of our models have present-day radii of 2-4 R$_\Earth$, consistent with the radius distribution of Kepler planet candidates, which exhibits a peak at 2-3 R$_\Earth$ \citep{2014ApJS..210...19B}. These radii are reached within $\sim$2 Gyr, with only slow radius evolution at later times.

Moreover, we find that the present-day radius of a planet is a very good proxy for its initial envelope mass fraction above a certain ``turn-off mass,'' below which mass loss dominates, and planets rapidly become ``terrestrial.'' This turn-off mass is dependent upon orbital distance and ranges from about 10 M$_\Earth$ at 0.05 AU to about 2 M$_\Earth$ at 0.30 AU, shifting by approximately 1-2 M$_\Earth$ for each factor of 2 in stellar irradiation. At higher masses, up to 20 M$_\Earth$, our mass-radius curves are approximately flat and mostly determined by their initial envelope mass fraction, a behavior that dominates from $f_{env,i} = 0.01$ to $f_{env,i} = 0.10$ and results in a spread of $\sim$2 R$_\Earth$ in radius at Gyr ages.

We also find that the initial radius of a planet has only a secondary effect on its subsequent evolution; there is little difference after 100 Myr between a close-in mini-Neptune planet formed with a relatively hot start and an initial radius of up to 20 R$_\Earth$ and one formed with a relatively cold start with an initial radius of 5 R$_\Earth$. Planets with initial radii of 10 and 20 R$_\Earth$ converge to nearly the same radius within just a few Myr, thus removing one important source of uncertainty in evolutionary modeling. The exception to this is the case of lower-mass planets for which mass loss dominates.

Lower-mass ($\sim$2 M$_\Earth$) planets in general have larger radii and experience more mass loss, as well as a greater spread in the amount of mass loss, with representative models losing 50\%-80\% of their envelopes, compared with higher-mass (4-8 M$_\Earth$) planets that lose 5\%-20\% of their envelopes (for representative models). In closer orbits, low-mass planets are also much more prone to losing their envelopes entirely.

The difference in semi-major axis is relatively small, a factor of 2-3, between a planet that retains most of its envelope and a planet that loses most or all of it, but this distance range varies with mass. For more massive planets of 16 M$_\Earth$, there is a transition from none of the envelope remaining at 2 Gyr at 0.03 AU to $\sim$85\% of it remaining at 0.10 AU. For low-mass planets of 2 M$_\Earth$, $\sim$3\% of the envelope remains at 2 Gyr at 0.20 AU, and a majority of it is retained only at $>$0.30 AU. A similar effect occurs in the radius near the inner boundaries of these ``transition regions'' as the loss of most of the envelope mass causes the radius to shrink rapidly.

Increasing metallicity results in a slight increase in the radius of a planet predominately due to the greater opacity and slower cooling of the envelope. At Gyr ages, this difference amounts to 0.4 R$_\Earth$ between 1x and 30x solar metallicity. The relatively small expected range of metallicities for mini-Neptune atmospheres ($\gtrsim$30x solar) means that metallicity is likely a small factor in variations in the real planet population.

The specific prescription for mass loss in our calculations is important. It is not possible to fit most of the low-mass, low-density close-in mini-Neptune planets without mass loss, and changing the mass loss efficiency and the time of the onset of mass loss (the migration time, $t_M$, which may be long after formation due to secular processes) can result in variations in radius of several tenths of an Earth radius at Gyr ages, with greater variations at low masses. There is significantly more variation in the amount of mass loss between models than in radius, but the envelope mass is a relatively small fraction of the total mass in our models, so the variation in mass loss does not result in a significant change in the expected mass distribution of planets over time.

In fitting observed planets in the Kepler-11 system, we find significantly lower envelope fractions than previous studies$-$by about a factor of 3 for Kepler-11d and -11e, for example. For Kepler-11d, we find a present-day $f_{env} = 2.6\pm 0.3\%$, compared with $f_{env} = 8.4_{-2.4}^{+2.7}\%$ computed by \citet{2012ApJ...761...59L}. Similarly, for Kepler-11e, we find $f_{env} = 6.5\pm 0.5\%$, compared with $17.2_{-4.2}^{+4.1}\%$ in the Lopez study. In addition to our reanalysis, we also achieve a preliminary fit to the latest mass and radius measurements of Kepler-11g of $f_{env} = 2.6\pm 0.3\%$. Further work is needed to refine these results through a better understanding of the equations of state, heat capacity, opacities, and other factors that contribute to super-Earth and sub-Neptune planetary evolution.

However, the lower-mass and lower-density planets of Kepler-11b, -11c, and -11f are challenging to model at their current locations. We successfully model Kepler-11f with $f_{env}\sim 0.0025$, an order of magnitude lower than the Lopez study. However, for Kepler-11b, we find no models that retain any H$_2$-He envelope, which is required to account for their radii, and for Kepler-11c, we find that either a very low mass loss efficiency or a late migration time is required to retain a H$_2$-He envelope.

For Kepler-11b, the only solution we find that is consistent with observations is that it is a ``water-world'' with an ice core and a thick steam atmosphere and no hydrogen or helium \citep{2012ApJ...756..176H,2014ApJ...787..173H}. While further study is needed on the evolution of a ``water-world'' in such an environment, we construct a plausible model to explain both Kepler-11b and Kepler-11c$-$likely our most plausible model for both planets$-$in which the two planets formed beyond the snow line and then migrated inwards to their current orbits due to secular processes at an age of $\sim$300 Myr (but not significantly earlier). Kepler-11b, being lower in mass, either formed without a H$_2$-He envelope or lost it entirely after migration, while Kepler-11c retains a small hydrogen envelope on the order of $f_{env}\sim 0.007$. Additional work is needed to investigate the possibility of extending this solution to all six planets in the Kepler-11 system.

Further study is required to determine the structures of very low-mass, low-density planets and to more accurately determine the effects of the properties of the atmospheres (e.g. anisotropic heating, thermal redistribution, and clouds) on super-Earth and mini-Neptune planetary evolution. More accurate mass measurements in particular are also needed for observed planets, given that the deduced compositions and evolutionary histories can be very sensitive to mass, to accurately fit structural and evolutionary models to these objects.

\begin{acknowledgements}

A.B. would like to acknowledge support in part under NASA grant NNX15AE19G, JPL subcontract no. 1513640, NASA HST awards HST-GO-13467.10-A, HST-GO-12550.02, and HST-GO-12473.06-A, and JPL/Spitzer contracts 1439064 and 1477493. We thank Eric Lopez and Ruth Murray-Clay for useful conversations and Lauren Weiss for her data on Kepler-11. We present a graphical user interface-based calculator demonstrating our structural model for the case of solid planets at http://www.astro.princeton.edu/$^\sim$arhowe/ and http://www.astro.princeton.edu/$^\sim$burrows/.

\end{acknowledgements}


\bibliographystyle{apj}
\bibliography{apj-jour,refs}

\begin{thebibliography}{39}
\expandafter\ifx\csname natexlab\endcsname\relax\def\natexlab#1{#1}\fi

\bibitem[{{Anderson} {et~al.}(2010){Anderson}, {Hellier}, {Gillon}, {Triaud},
  {Smalley}, {Hebb}, {Collier Cameron}, {Maxted}, {Queloz}, {West}, {Bentley},
  {Enoch}, {Horne}, {Lister}, {Mayor}, {Parley}, {Pepe}, {Pollacco},
  {S{\'e}gransan}, {Udry}, \& {Wilson}}]{2010ApJ...709..159A}
{Anderson}, D.~R., {et~al.} 2010, \apj, 709, 159

\bibitem[{{Baraffe} {et~al.}(2004){Baraffe}, {Selsis}, {Chabrier}, {Barman},
  {Allard}, {Hauschildt}, \& {Lammer}}]{2004A&A...419L..13B}
{Baraffe}, I., {Selsis}, F., {Chabrier}, G., {Barman}, T.~S., {Allard}, F.,
  {Hauschildt}, P.~H., \& {Lammer}, H. 2004, \aap, 419, L13

\bibitem[{{Burke} {et~al.}(2014){Burke}, {Bryson}, {Mullally}, {Rowe},
  {Christiansen}, {Thompson}, {Coughlin}, {Haas}, {Batalha}, {Caldwell},
  {Jenkins}, {Still}, {Barclay}, {Borucki}, {Chaplin}, {Ciardi}, {Clarke},
  {Cochran}, {Demory}, {Esquerdo}, {Gautier}, {Gilliland}, {Girouard}, {Havel},
  {Henze}, {Howell}, {Huber}, {Latham}, {Li}, {Morehead}, {Morton}, {Pepper},
  {Quintana}, {Ragozzine}, {Seader}, {Shah}, {Shporer}, {Tenenbaum}, {Twicken},
  \& {Wolfgang}}]{2014ApJS..210...19B}
{Burke}, C.~J., {et~al.} 2014, \apjs, 210, 19

\bibitem[{{Burrows} {et~al.}(2006){Burrows}, {Sudarsky}, \&
  {Hubeny}}]{2006ApJ...650.1140B}
{Burrows}, A., {Sudarsky}, D., \& {Hubeny}, I. 2006, \apj, 650, 1140

\bibitem[{{Cochran} {et~al.}(2011){Cochran}, {Fabrycky}, {Torres}, {Fressin},
  {D{\'e}sert}, {Ragozzine}, {Sasselov}, {Fortney}, {Rowe}, {Brugamyer},
  {Bryson}, {Carter}, {Ciardi}, {Howell}, {Steffen}, {Borucki}, {Koch}, {Winn},
  {Welsh}, {Uddin}, {Tenenbaum}, {Still}, {Seager}, {Quinn}, {Mullally},
  {Miller}, {Marcy}, {MacQueen}, {Lucas}, {Lissauer}, {Latham}, {Knutson},
  {Kinemuchi}, {Johnson}, {Jenkins}, {Isaacson}, {Howard}, {Horch}, {Holman},
  {Henze}, {Haas}, {Gilliland}, {Gautier}, {Ford}, {Fischer}, {Everett},
  {Endl}, {Demory}, {Deming}, {Charbonneau}, {Caldwell}, {Buchhave}, {Brown},
  \& {Batalha}}]{2011ApJS..197....7C}
{Cochran}, W.~D., {et~al.} 2011, \apjs, 197, 7

\bibitem[{{Dressing} {et~al.}(2015){Dressing}, {Charbonneau}, {Dumusque},
  {Gettel}, {Pepe}, {Collier Cameron}, {Latham}, {Molinari}, {Udry}, {Affer},
  {Bonomo}, {Buchhave}, {Cosentino}, {Figueira}, {Fiorenzano}, {Harutyunyan},
  {Haywood}, {Johnson}, {Lopez-Morales}, {Lovis}, {Malavolta}, {Mayor},
  {Micela}, {Motalebi}, {Nascimbeni}, {Phillips}, {Piotto}, {Pollacco},
  {Queloz}, {Rice}, {Sasselov}, {S{\'e}gransan}, {Sozzetti}, {Szentgyorgyi}, \&
  {Watson}}]{2015ApJ...800..135D}
{Dressing}, C.~D., {et~al.} 2015, \apj, 800, 135

\bibitem[{{Ehrenreich} {et~al.}(2008){Ehrenreich}, {Lecavelier Des Etangs},
  {H{\'e}brard}, {D{\'e}sert}, {Vidal-Madjar}, {McConnell}, {Parkinson},
  {Ballester}, \& {Ferlet}}]{2008A&A...483..933E}
{Ehrenreich}, D., {et~al.} 2008, \aap, 483, 933

\bibitem[{{Helled} {et~al.}(2011){Helled}, {Anderson}, {Podolak}, \&
  {Schubert}}]{2011ApJ...726...15H}
{Helled}, R., {Anderson}, J.~D., {Podolak}, M., \& {Schubert}, G. 2011, \apj,
  726, 15

\bibitem[{{Howe} {et~al.}(2014){Howe}, {Burrows}, \&
  {Verne}}]{2014ApJ...787..173H}
{Howe}, A.~R., {Burrows}, A., \& {Verne}, W. 2014, \apj, 787, 173

\bibitem[{{Howe} \& {Burrows}(2012)}]{2012ApJ...756..176H}
{Howe}, A.~R., \& {Burrows}, A.~S. 2012, \apj, 756, 176

\bibitem[{{Hubbard} {et~al.}(2007){Hubbard}, {Hattori}, {Burrows}, {Hubeny}, \&
  {Sudarsky}}]{2007Icar..187..358H}
{Hubbard}, W.~B., {Hattori}, M.~F., {Burrows}, A., {Hubeny}, I., \& {Sudarsky},
  D. 2007, Icarus, 187, 358

\bibitem[{{Hubeny} {et~al.}(2003){Hubeny}, {Burrows}, \&
  {Sudarsky}}]{2003ApJ...594.1011H}
{Hubeny}, I., {Burrows}, A., \& {Sudarsky}, D. 2003, \apj, 594, 1011

\bibitem[{{Jin} {et~al.}(2014){Jin}, {Mordasini}, {Parmentier}, {van Boekel},
  {Henning}, \& {Ji}}]{2014ApJ...795...65J}
{Jin}, S., {Mordasini}, C., {Parmentier}, V., {van Boekel}, R., {Henning}, T.,
  \& {Ji}, J. 2014, \apj, 795, 65

\bibitem[{{Jontof-Hutter} {et~al.}(2014){Jontof-Hutter}, {Lissauer}, {Rowe}, \&
  {Fabrycky}}]{2014ApJ...785...15J}
{Jontof-Hutter}, D., {Lissauer}, J.~J., {Rowe}, J.~F., \& {Fabrycky}, D.~C.
  2014, \apj, 785, 15

\bibitem[{{Lecavelier Des Etangs} {et~al.}(2008){Lecavelier Des Etangs},
  {Pont}, {Vidal-Madjar}, \& {Sing}}]{2008A&A...481L..83L}
{Lecavelier Des Etangs}, A., {Pont}, F., {Vidal-Madjar}, A., \& {Sing}, D.
  2008, \aap, 481, L83

\bibitem[{{Lee} {et~al.}(2014){Lee}, {Chiang}, \&
  {Ormel}}]{2014ApJ...797...95L}
{Lee}, E.~J., {Chiang}, E., \& {Ormel}, C.~W. 2014, \apj, 797, 95

\bibitem[{{Lissauer} {et~al.}(2011){Lissauer}, {Fabrycky}, {Ford}, {Borucki},
  {Fressin}, {Marcy}, {Orosz}, {Rowe}, {Torres}, {Welsh}, {Batalha}, {Bryson},
  {Buchhave}, {Caldwell}, {Carter}, {Charbonneau}, {Christiansen}, {Cochran},
  {Desert}, {Dunham}, {Fanelli}, {Fortney}, {Gautier}, {Geary}, {Gilliland},
  {Haas}, {Hall}, {Holman}, {Koch}, {Latham}, {Lopez}, {McCauliff}, {Miller},
  {Morehead}, {Quintana}, {Ragozzine}, {Sasselov}, {Short}, \&
  {Steffen}}]{2011Natur.470...53L}
{Lissauer}, J.~J., {et~al.} 2011, Nature, 470, 53

\bibitem[{{Lissauer} {et~al.}(2013){Lissauer}, {Jontof-Hutter}, {Rowe},
  {Fabrycky}, {Lopez}, {Agol}, {Marcy}, {Deck}, {Fischer}, {Fortney}, {Howell},
  {Isaacson}, {Jenkins}, {Kolbl}, {Sasselov}, {Short}, \&
  {Welsh}}]{2013ApJ...770..131L}
---. 2013, \apj, 770, 131

\bibitem[{{Lopez} \& {Fortney}(2013)}]{2013ApJ...776....2L}
{Lopez}, E.~D., \& {Fortney}, J.~J. 2013, \apj, 776, 2

\bibitem[{{Lopez} \& {Fortney}(2014)}]{2014ApJ...792....1L}
---. 2014, \apj, 792, 1

\bibitem[{{Lopez} {et~al.}(2012){Lopez}, {Fortney}, \&
  {Miller}}]{2012ApJ...761...59L}
{Lopez}, E.~D., {Fortney}, J.~J., \& {Miller}, N. 2012, \apj, 761, 59

\bibitem[{{Marcy} {et~al.}(2014){Marcy}, {Isaacson}, {Howard}, {Rowe},
  {Jenkins}, {Bryson}, {Latham}, {Howell}, {Gautier}, {Batalha}, {Rogers},
  {Ciardi}, {Fischer}, {Gilliland}, {Kjeldsen}, {Christensen-Dalsgaard},
  {Huber}, {Chaplin}, {Basu}, {Buchhave}, {Quinn}, {Borucki}, {Koch}, {Hunter},
  {Caldwell}, {Van Cleve}, {Kolbl}, {Weiss}, {Petigura}, {Seager}, {Morton},
  {Johnson}, {Ballard}, {Burke}, {Cochran}, {Endl}, {MacQueen}, {Everett},
  {Lissauer}, {Ford}, {Torres}, {Fressin}, {Brown}, {Steffen}, {Charbonneau},
  {Basri}, {Sasselov}, {Winn}, {Sanchis-Ojeda}, {Christiansen}, {Adams},
  {Henze}, {Dupree}, {Fabrycky}, {Fortney}, {Tarter}, {Holman}, {Tenenbaum},
  {Shporer}, {Lucas}, {Welsh}, {Orosz}, {Bedding}, {Campante}, {Davies},
  {Elsworth}, {Handberg}, {Hekker}, {Karoff}, {Kawaler}, {Lund}, {Lundkvist},
  {Metcalfe}, {Miglio}, {Silva Aguirre}, {Stello}, {White}, {Boss}, {Devore},
  {Gould}, {Prsa}, {Agol}, {Barclay}, {Coughlin}, {Brugamyer}, {Mullally},
  {Quintana}, {Still}, {Thompson}, {Morrison}, {Twicken}, {D{\'e}sert},
  {Carter}, {Crepp}, {H{\'e}brard}, {Santerne}, {Moutou}, {Sobeck}, {Hudgins},
  {Haas}, {Robertson}, {Lillo-Box}, \& {Barrado}}]{2014ApJS..210...20M}
{Marcy}, G.~W., {et~al.} 2014, \apjs, 210, 20

\bibitem[{{Masuda}(2014)}]{2014ApJ...783...53M}
{Masuda}, K. 2014, \apj, 783, 53

\bibitem[{{Masuda} {et~al.}(2013){Masuda}, {Hirano}, {Taruya}, {Nagasawa}, \&
  {Suto}}]{2013ApJ...778..185M}
{Masuda}, K., {Hirano}, T., {Taruya}, A., {Nagasawa}, M., \& {Suto}, Y. 2013,
  \apj, 778, 185

\bibitem[{{Mordasini} {et~al.}(2012{\natexlab{a}}){Mordasini}, {Alibert},
  {Georgy}, {Dittkrist}, {Klahr}, \& {Henning}}]{2012A&A...547A.112M}
{Mordasini}, C., {Alibert}, Y., {Georgy}, C., {Dittkrist}, K.-M., {Klahr}, H.,
  \& {Henning}, T. 2012{\natexlab{a}}, \aap, 547, A112

\bibitem[{{Mordasini} {et~al.}(2012{\natexlab{b}}){Mordasini}, {Alibert},
  {Klahr}, \& {Henning}}]{2012A&A...547A.111M}
{Mordasini}, C., {Alibert}, Y., {Klahr}, H., \& {Henning}, T.
  2012{\natexlab{b}}, \aap, 547, A111

\bibitem[{{Murray-Clay} {et~al.}(2009){Murray-Clay}, {Chiang}, \&
  {Murray}}]{2009ApJ...693...23M}
{Murray-Clay}, R.~A., {Chiang}, E.~I., \& {Murray}, N. 2009, \apj, 693, 23

\bibitem[{{Nettelmann} {et~al.}(2011){Nettelmann}, {Fortney}, {Kramm}, \&
  {Redmer}}]{2011ApJ...733....2N}
{Nettelmann}, N., {Fortney}, J.~J., {Kramm}, U., \& {Redmer}, R. 2011, \apj,
  733, 2

\bibitem[{{Ofir} {et~al.}(2014){Ofir}, {Dreizler}, {Zechmeister}, \&
  {Husser}}]{2014A&A...561A.103O}
{Ofir}, A., {Dreizler}, S., {Zechmeister}, M., \& {Husser}, T.-O. 2014, \aap,
  561, A103

\bibitem[{{Owen} \& {Wu}(2013)}]{2013ApJ...775..105O}
{Owen}, J.~E., \& {Wu}, Y. 2013, \apj, 775, 105

\bibitem[{{Rafikov}(2006)}]{2006ApJ...648..666R}
{Rafikov}, R.~R. 2006, \apj, 648, 666

\bibitem[{{Ribas} {et~al.}(2005){Ribas}, {Guinan}, {G{\"u}del}, \&
  {Audard}}]{2005ApJ...622..680R}
{Ribas}, I., {Guinan}, E.~F., {G{\"u}del}, M., \& {Audard}, M. 2005, \apj, 622,
  680

\bibitem[{{Sanz-Forcada} {et~al.}(2010){Sanz-Forcada}, {Ribas}, {Micela},
  {Pollock}, {Garc{\'{\i}}a-{\'A}lvarez}, {Solano}, \&
  {Eiroa}}]{2010A&A...511L...8S}
{Sanz-Forcada}, J., {Ribas}, I., {Micela}, G., {Pollock}, A.~M.~T.,
  {Garc{\'{\i}}a-{\'A}lvarez}, D., {Solano}, E., \& {Eiroa}, C. 2010, \aap,
  511, L8

\bibitem[{{Shkolnik} \& {Barman}(2014)}]{2014AJ....148...64S}
{Shkolnik}, E.~L., \& {Barman}, T.~S. 2014, \aj, 148, 64

\bibitem[{{Silburt} {et~al.}(2014){Silburt}, {Gaidos}, \&
  {Wu}}]{2014arXiv1406.6048S}
{Silburt}, A., {Gaidos}, E., \& {Wu}, Y. 2014, ArXiv e-prints

\bibitem[{{Watson} {et~al.}(1981){Watson}, {Donahue}, \&
  {Walker}}]{1981Icar...48..150W}
{Watson}, A.~J., {Donahue}, T.~M., \& {Walker}, J.~C.~G. 1981, Icarus, 48, 150

\bibitem[{{Weiss} \& {Marcy}(2014)}]{2014ApJ...783L...6W}
{Weiss}, L.~M., \& {Marcy}, G.~W. 2014, \apjl, 783, L6

\bibitem[{{Weiss} {et~al.}(2015){Weiss}, {Marcy}, {Isaacson}, {Fulton}, {Rowe},
  \& {Lissauer}}]{2015Weiss..in.press}
{Weiss}, L.~M., {Marcy}, G.~W., {Isaacson}, K.~D., {Fulton}, B.~J., {Rowe}, J.,
  \& {Lissauer}, D.~J.-H. 2015, in prep.

\bibitem[{{Wolfgang} \& {Lopez}(2014)}]{2014arXiv1409.2982W}
{Wolfgang}, A., \& {Lopez}, E. 2014, ArXiv e-prints

\end{thebibliography}


\end{document}